\documentstyle{mn}

\newif\ifAMStwofonts

\ifoldfss
  \ifCUPmtlplainloaded \else
    \NewTextAlphabet{textbfit} {cmbxti10} {}
    \NewTextAlphabet{textbfss} {cmssbx10} {}
    \NewMathAlphabet{mathbfit} {cmbxti10} {} 
    \NewMathAlphabet{mathbfss} {cmssbx10} {} 
  \fi
  \ifAMStwofonts
    \ifCUPmtlplainloaded \else
      \NewSymbolFont{upmath} {eurm10}
      \NewSymbolFont{AMSa} {msam10}
      \NewMathSymbol{\upi}     {0}{upmath}{19}
      \NewMathSymbol{\umu}     {0}{upmath}{16}
      \NewMathSymbol{\upartial}{0}{upmath}{40}
      \NewMathSymbol{\leqslant}{3}{AMSa}{36}
      \NewMathSymbol{\geqslant}{3}{AMSa}{3E}

    \fi
  \fi
\fi 

\ifnfssone
  \newmathalphabet{\mathit}
  \addtoversion{normal}{\mathit}{cmr}{m}{it}
  \addtoversion{bold}{\mathit}{cmr}{bx}{it}
  \newmathalphabet{\mathbfit} 
  \addtoversion{normal}{\mathbfit}{cmr}{bx}{it}
  \addtoversion{bold}{\mathbfit}{cmr}{bx}{it}
  \newmathalphabet{\mathbfss} 
  \addtoversion{normal}{\mathbfss}{cmss}{bx}{n}
  \addtoversion{bold}{\mathbfss}{cmss}{bx}{n}
  \ifAMStwofonts
    \ifCUPmtlplainloaded \else
      \UseAMStwoboldmath
      \makeatletter
      \new@mathgroup\upmath@group
      \define@mathgroup\mv@normal\upmath@group{eur}{m}{n}
      \define@mathgroup\mv@bold\upmath@group{eur}{b}{n}
      \edef\UPM{\hexnumber\upmath@group}
      \new@mathgroup\amsa@group
      \define@mathgroup\mv@normal\amsa@group{msa}{m}{n}
      \define@mathgroup\mv@bold\amsa@group{msa}{m}{n}
      \edef\AMSa{\hexnumber\amsa@group}
      \makeatother
      \mathchardef\upi="0\UPM19
      \mathchardef\umu="0\UPM16
      \mathchardef\upartial="0\UPM40
      \mathchardef\leqslant="3\AMSa36
      \mathchardef\geqslant="3\AMSa3E
    \fi
  \fi
\fi 

\ifnfsstwo
  \DeclareMathAlphabet{\mathbfit}{OT1}{cmr}{bx}{it}
  \SetMathAlphabet\mathbfit{bold}{OT1}{cmr}{bx}{it}
  \DeclareMathAlphabet{\mathbfss}{OT1}{cmss}{bx}{n}
  \SetMathAlphabet\mathbfss{bold}{OT1}{cmss}{bx}{n}
  \ifAMStwofonts
    \ifCUPmtlplainloaded \else
      \DeclareSymbolFont{UPM}{U}{eur}{m}{n}
      \SetSymbolFont{UPM}{bold}{U}{eur}{b}{n}
      \DeclareSymbolFont{AMSa}{U}{msa}{m}{n}
      \DeclareMathSymbol{\upi}{0}{UPM}{"19}
      \DeclareMathSymbol{\umu}{0}{UPM}{"16}
      \DeclareMathSymbol{\upartial}{0}{UPM}{"40}
      \DeclareMathSymbol{\leqslant}{3}{AMSa}{"36}
      \DeclareMathSymbol{\geqslant}{3}{AMSa}{"3E}
    \fi
  \fi
\fi 

\ifCUPmtlplainloaded \else
  \ifAMStwofonts \else 
    \def\upi{\pi}
    \def\umu{\mu}
    \def\upartial{\partial}
  \fi
\fi

\title{Changes in the measured image separation of the gravitational
  lens system, PKS~1830-211.} 

\author[C. Jin et al.]  {C.~Jin,$^1$ M.A.~Garrett,$^2$
  S.~Nair,$^3$ R.W.~Porcas,$^4$
  A.R.~Patnaik,$^4$ R.~Nan$^1$ \\
  $^1$National Astronomical Observatories, Chinese Academy of
  Sciences, A20 Datun Road, Chaoyang District, Beijing 100012, China\\
  $^2$Joint Institute for VLBI in Europe, Postbus 2,
  7990~AA Dwingeloo, The Netherlands \\
  $^3$Raman Research Institute, C.V. Raman Avenue,
  Bangalore 560080, India \\
  $^4$Max-Planck-Institut f\"ur Radioastronomie, Auf dem H\"ugel
  69,D-53121 Bonn,Germany }
         
\date{Accepted 2003 January 8. 
      Received 2002 March 13;
      in original form 2002 March 7}

\pagerange{\pageref{firstpage}--\pageref{lastpage}}
\pubyear{2001}

\begin{document}

\maketitle

\label{firstpage}

\begin{abstract}
  
  We present eight epochs of 43~GHz, dual-polarisation VLBA
  observations of the gravitational lens system PKS 1830-211, made over
  fourteen weeks. A bright, compact
  ``core'' and a faint extended ``jet'' are clearly seen in maps of
  both lensed images at all eight epochs.  The relative separation of
  the radio centroid of the cores (as measured on the sky) changes by
  up to 87 $\mu$as between subsequent epochs. 
  A comparison with the previous 43~GHz VLBA
  observations \cite{b6} made 8 months earlier show even larger 
  deviations in the separation of up to 201 $\mu$as. 
  The measured
  changes are most likely produced by changes in the brightness
  distribution of the background source, enhanced by the magnification
  of the lens.
  A relative magnification matrix that is
  applicable on the milliarcsecond scale has been determined by
  relating two vectors (the ``core-jet'' separations and the offsets of
  the polarised and total intensity emission) in the two lensed images.
  The determinant of this matrix, $-1.13$ $(\pm 0.61)$, is in good
  agreement with the measured flux density ratio of the two images. 
  The matrix predicts that the 10~mas long jet, that is clearly seen in
  previous 15 and 8.4~GHz VLBA observations 
(Garrett et al. 1997, Guirado et al. 1999),
  should correspond to a 4 mas long jet trailing to the south-east of
  the SW image. The clear non-detection of this trailing jet 
  is a strong evidence for sub-structure in the lens and  may 
  require more realistic lens models to be invoked 
  {\it e.g.} Nair \& Garrett \shortcite{b14}.

\end{abstract}


\section{Introduction}

PKS 1830-211 is a very bright and highly variable radio source at cm-
and mm-wavelengths \cite{b17}. As well as being a double-imaged
gravitational lens system 
(Rao \& Subrahmanyan 1988, Jauncey et al. 1991),
it is also \\identified
by the EGRET instrument on the Compton Gamma-ray Observatory as an 
unusually strong source of gamma-rays
\cite{b5}. The radio and gamma-ray properties suggest that the
background source in this system is best classified as a blazar
\cite{b2}. The source lies close to the galactic plane, and optical
identifications and redshift determinations have been severely
handicapped by the high level of dust obscuration along the
line-of-sight. However, 
optical and infrared observations at Keck and ESO 
(Courbin et al. 1998, Frye et al. 1999)
and infrared observation using HST \cite{b24} have identified both of 
the flat spectrum radio cores. 
Recently, the ESO \\
New Technology Telescope(NTT) near infra-red spectra 
show clear detections of both the $H_\alpha$ and $H_\beta$ 
emission lines, implying a source redshift of $z_s = 2.507$ \cite{b9}.
The lens may be a compound system: molecular \\absorption at 
$z=0.886$ \cite{b18} and HI absorption at $z=0.19$ \cite{b19} have 
been \\detected.
A time delay ($26^{+4}_{-5}$ days \cite{b10}; $24^{+5}_{-4}$ days 
\cite{b26}), and a magnification ratio of 
$1.52 \pm 0.005$ \cite{b10} have been measured for the two compact core
images in the system.

Previous high resolution radio observations of the source show two lensed 
images, with a separation of about 1 arcsec \cite{b4}. MERLIN and
VLA observations revealed an elliptical Einstein ring connecting the
two brighter components \cite{b15}. \\Subsequent VLBI observations
(Garrett et al. 1997, Guirado et al. 1999)
have revealed detailed structures in both of the
two lensed images. VLBI observations at 8.4~GHz \\ 
\cite{b16}, 15~GHz \cite{b6}
and 22~GHz \cite{b13}, show significant differences between the
north-eastern (NE) and south-western (SW) images 
on the milliarcsecond (mas) scale. In particular, a prominent
10-mas-long jet associated with the core of the NE image, has no
obvious counterpart in the SW image. Nair \& Garrett \shortcite{b14}
attempt to explain this results as due to a perturbation in the NE
image by a lens of globular cluster scale. The first 43~GHz maps also
showed relatively \\
rapid changes in the brightness distribution of the
images \cite{b6}, an effect that may be partly explained by the
magnification provided by the lens system.  According \\
to models of this
system, the magnification between the images and background source, may
be as large as 5 - 10 
(Kochanek \& Narayan 1992, Nair et al. 1993).
 
In this paper, we present 8-epoch, VLBA 43~GHz maps \\
of both lensed
radio images, in both polarised and total\\
 intensity. In the following
section, a brief summary of the observations and data analysis are
presented. Measured changes in separation between the radio core images
are \\
presented and discussed in section 3, together with the \\
determination of a magnification matrix that relates the two images on
the milliarcsecond scale. A summary of the main results is 
presented in section 4.

\section[]{Observation and data reduction}

PKS~1830-211 was observed using the VLBA at 43~GHz at eight separate epochs,
over a period of 14 weeks (1997 \\ 
January 19 - 1997 April 30). The source was observed for a total of 5 hours at 
each epoch. And each epoch was \\separated in time by about 14 days. 
For all observations right and left hands of polarisation were
recorded in 2-bit mode, each with a total bandwidth of 32 MHz, divided 
\\into 4 IF channels. Several
nearby and compact calibrators (B~1730-130, B~1749+096, B~1741-038, and
B~1908-211) were observed every 18 minutes - at the beginning and end
of each 22 minute tape pass. Both parallel and cross-hand polarisation products
were generated by the NRAO VLBA correlator in Socorro, NM, USA. Only
one processing centre was \\
employed, positioned mid-way between the NE
and SW \\
images.  In order to avoid the effects of time smearing, very
short integrations times were provided by the correlator (0.39 s).

The data quality was generally
good except for the sixth epoch (1997 April 3) which was adversely
affected by bad weather at Kitt Peak, excessive noise and resulting
weak fringes at Brewster and bad playback at Fort Davis (for some but
not all IF channels). Nevertheless, the data for the sixth epoch were 
still reasonable after appropriate \\
editing of the data. 
Initial calibration of the data was \\
performed with the {\sc AIPS}
package.  The visibility amplitudes were calibrated using the system
temperatures and gain \\
information provided by each telescope.  Residual
delays and fringe rates associated with instrumental effects were \\
determined from the calibrators and removed from the \\
target source data
(PKS~1830-211).  Hybrid mapping of the calibrators also generated
antenna based amplitude gain \\corrections and these were also applied
and interpolated to PKS~1830-211. The instrumental polarisation calibration was
derived from the compact calibrators (following \\Lepp\"{a}nen et al.,
1995) and applied to PKS~1830-211. The polarization angle was not 
calibrated, since we were more \\interested in a comparison of the polarization
intensity and angle between the two lensed images which didn't require any 
absolute angle to be determined. At all eight epochs,
the D-terms (so-called {\it leakage factors}) solved are about \\
4-8~\% for BR, HN and LA and less than 5~\% for the other seven antennas in the 
VLBA array.

Large residual rates associated with atmospheric \\
instabilities remained
in the target data and this \\
necessitated fringe-fitting PKS~1830-211
directly. Due to the complicated nature of the source, it was first
necessary to produce an initial model of the source (containing both \\
images) using the (calibrator) corrected data. Preliminary total
intensity HYBRID maps of both lensed images were made using wide-field
techniques \cite{b1}, and these were later used as a model for
fringe-fitting. After this \\
final round of fringe-fitting the
corrections were applied to PKS~1830-211.

In order to reduce the size of the calibrated data set, each of the
individual IFs were averaged in frequency - the final data set being
composed of 4 independent (but \\
contiguous) 8 MHz channels (for each
parallel and cross hand polarisation product). The calibrated data were
also averaged in time in a baseline dependent manner \\
(integration times
ranged from 0.39 to 15 seconds depending on the projected baseline
length). Hybrid maps of both lensed images were made simultaneously from
this data set.  

\begin{figure*}
\vspace{19cm}
\begin{picture}(40,40)
\put(-240,0){\includegraphics{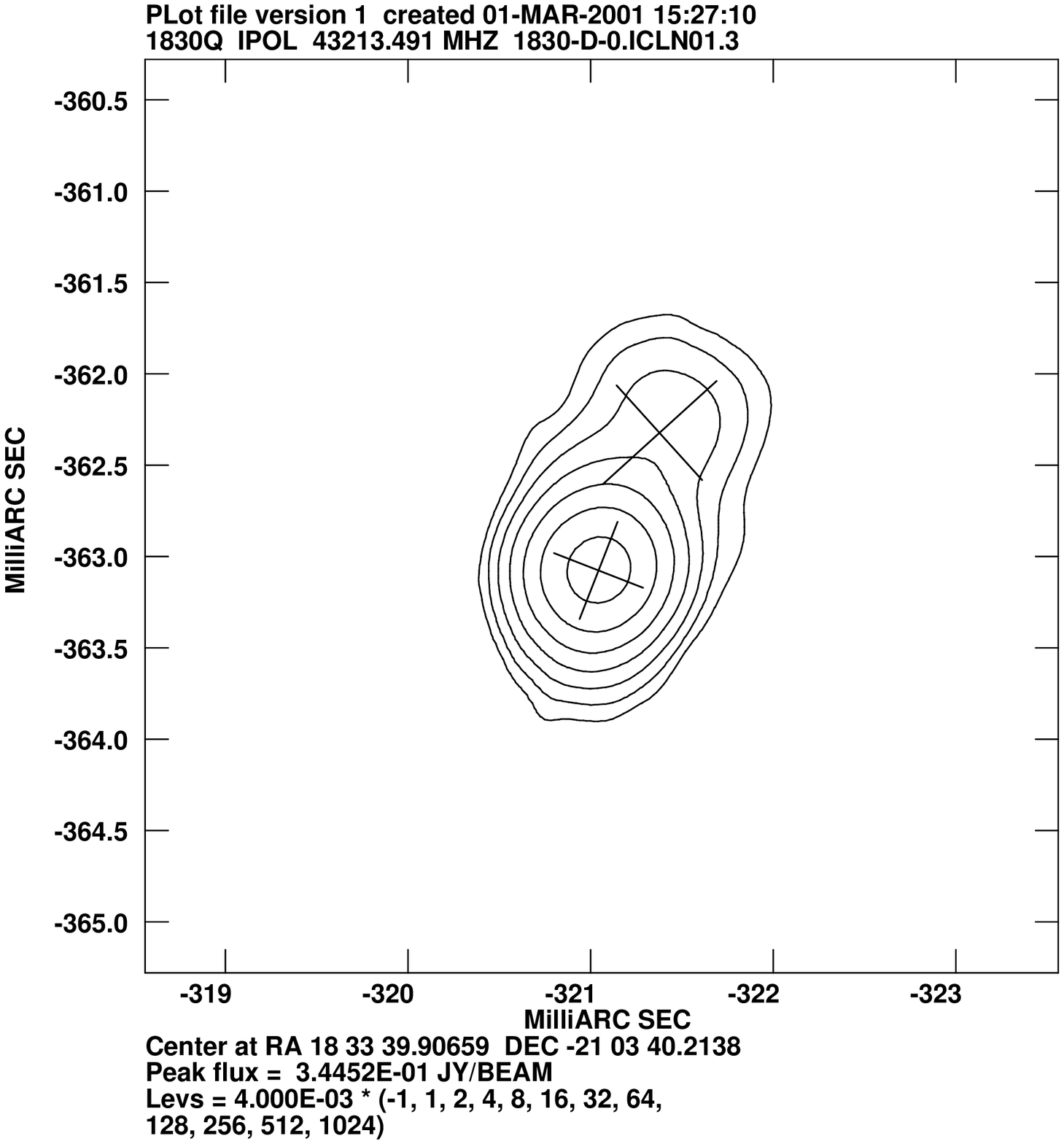}}
\put(-130,0){\includegraphics{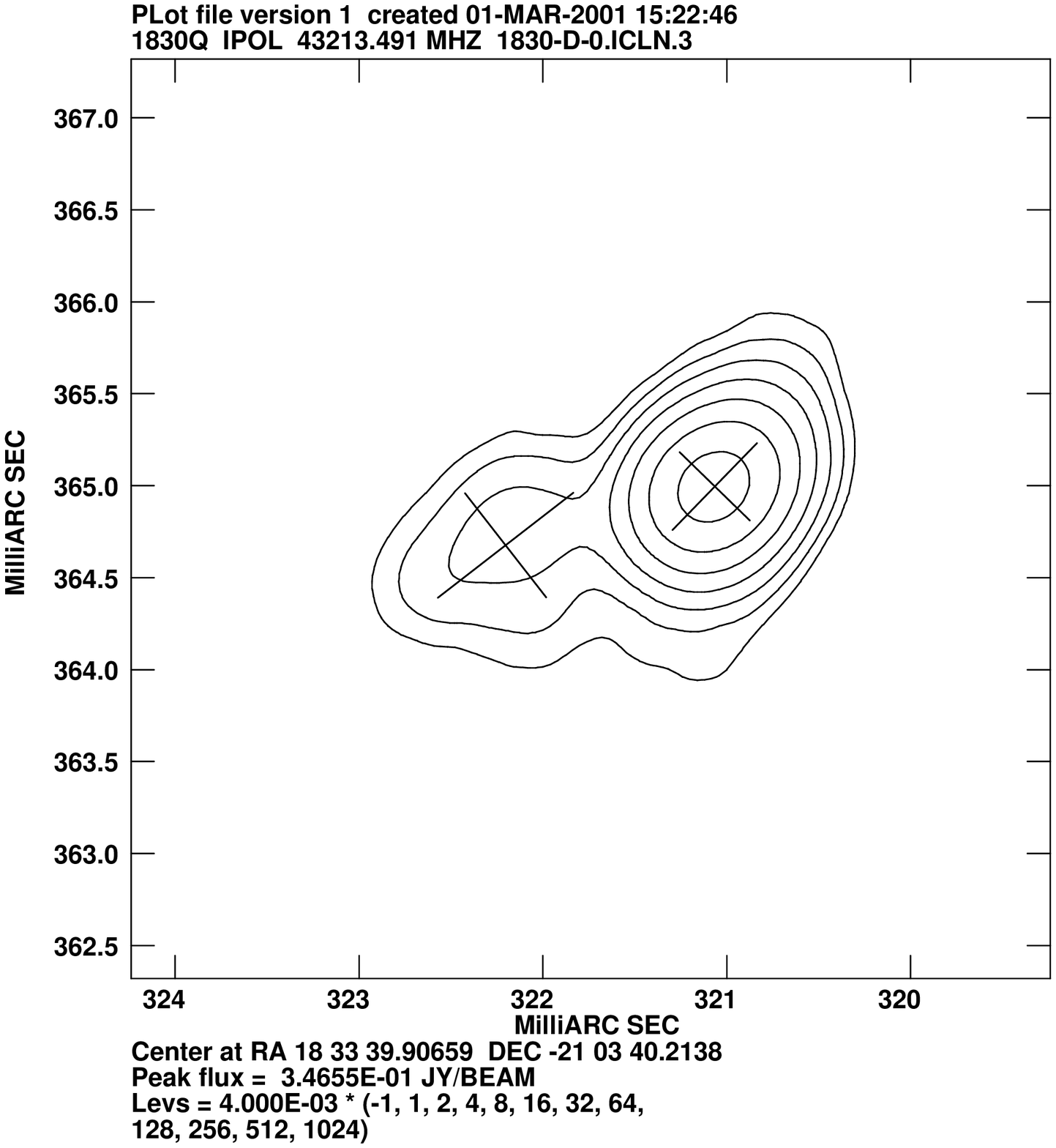}}
\put(-240,130){\includegraphics{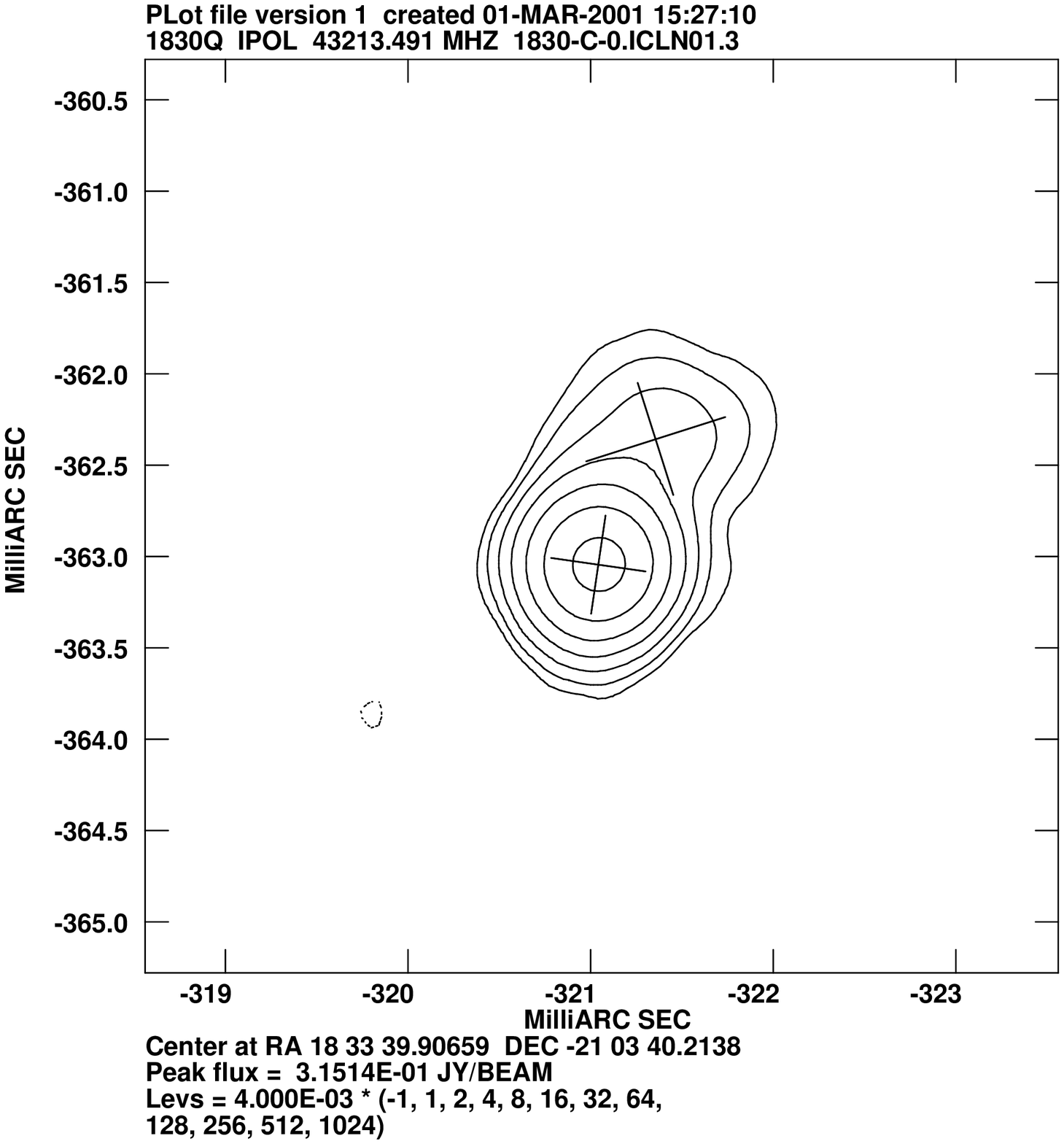}}
\put(-130,130){\includegraphics{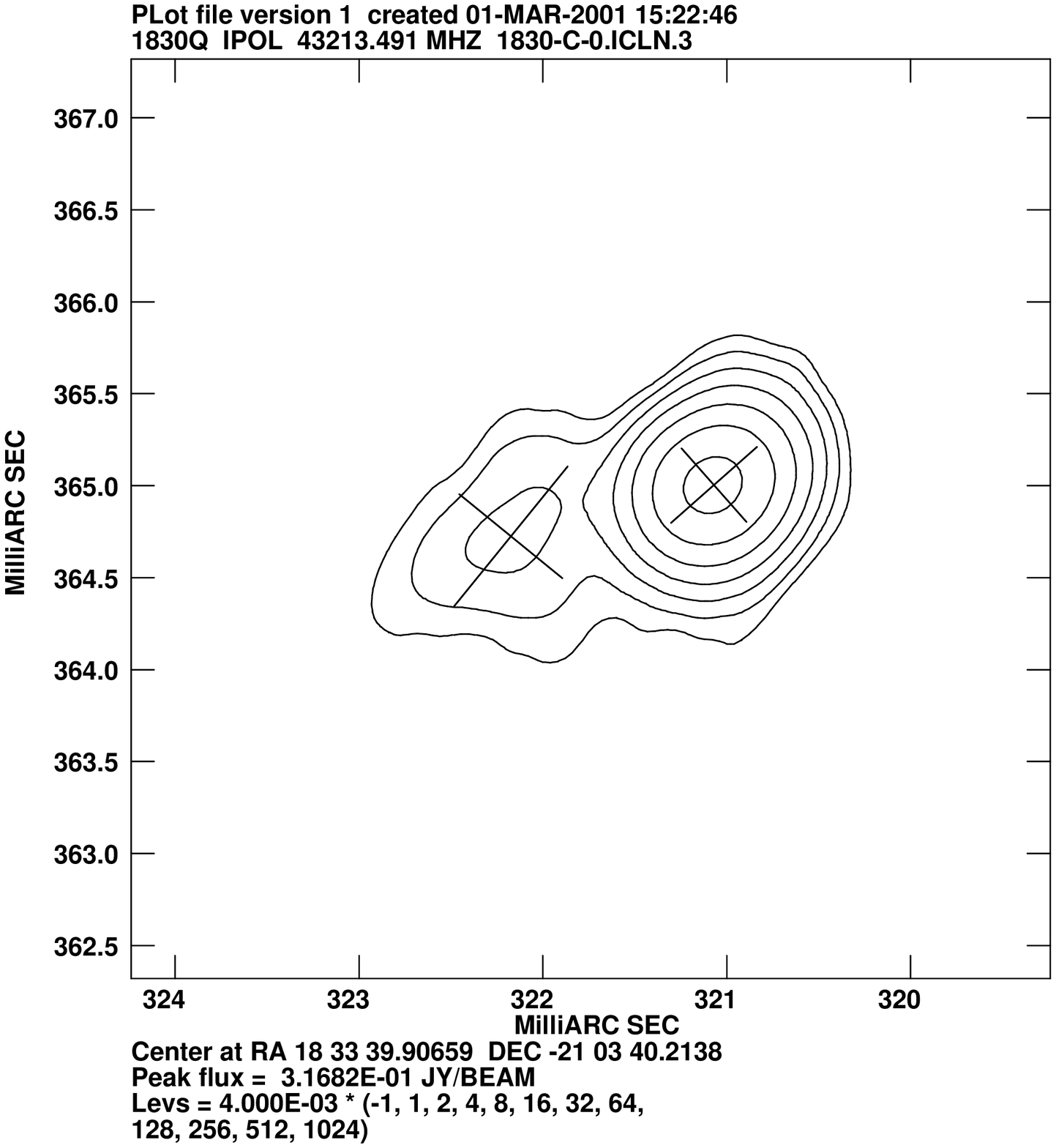}}
\put(-240,260){\includegraphics{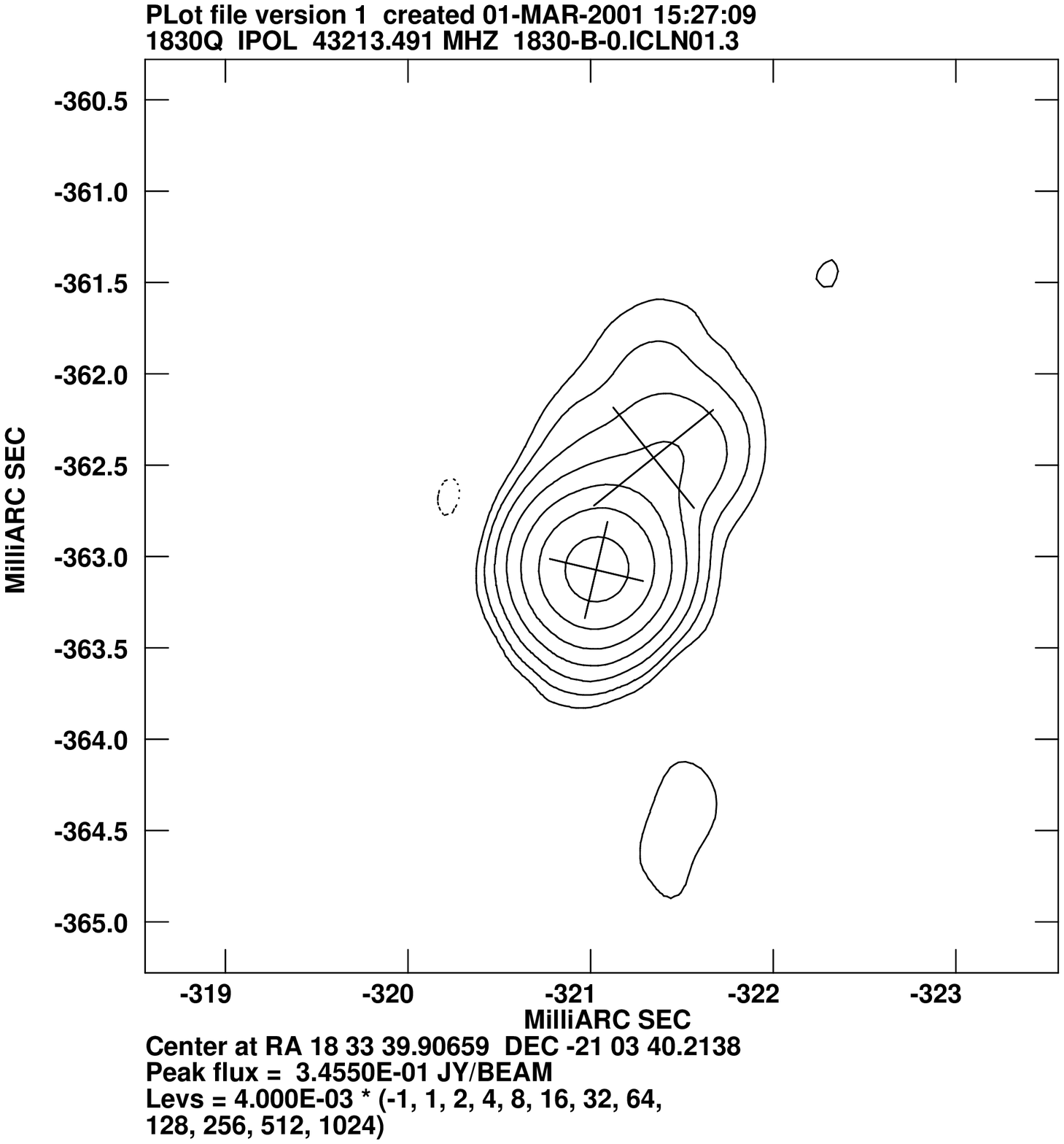}}
\put(-130,260){\includegraphics{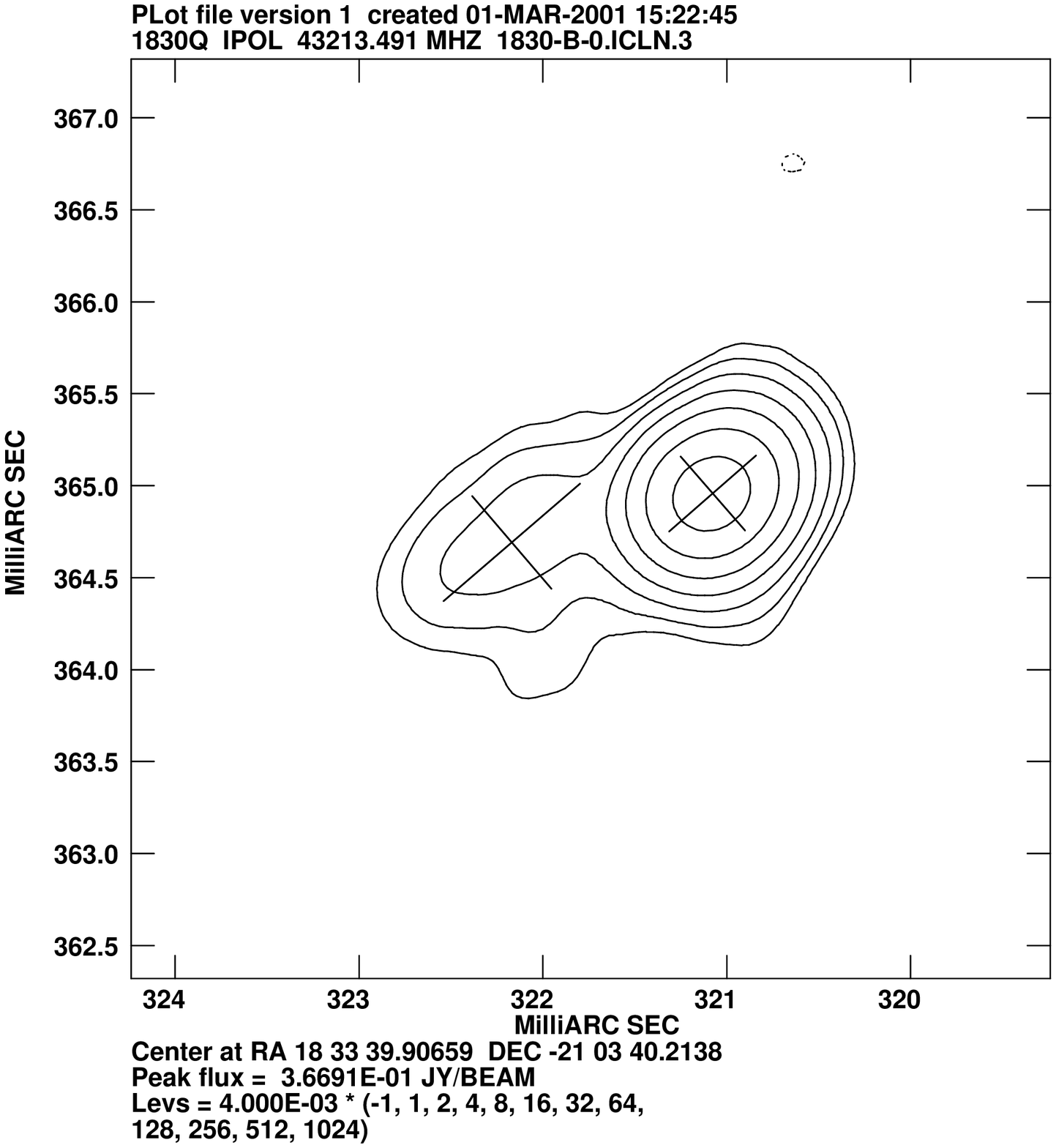}}
\put(-240,390){\includegraphics{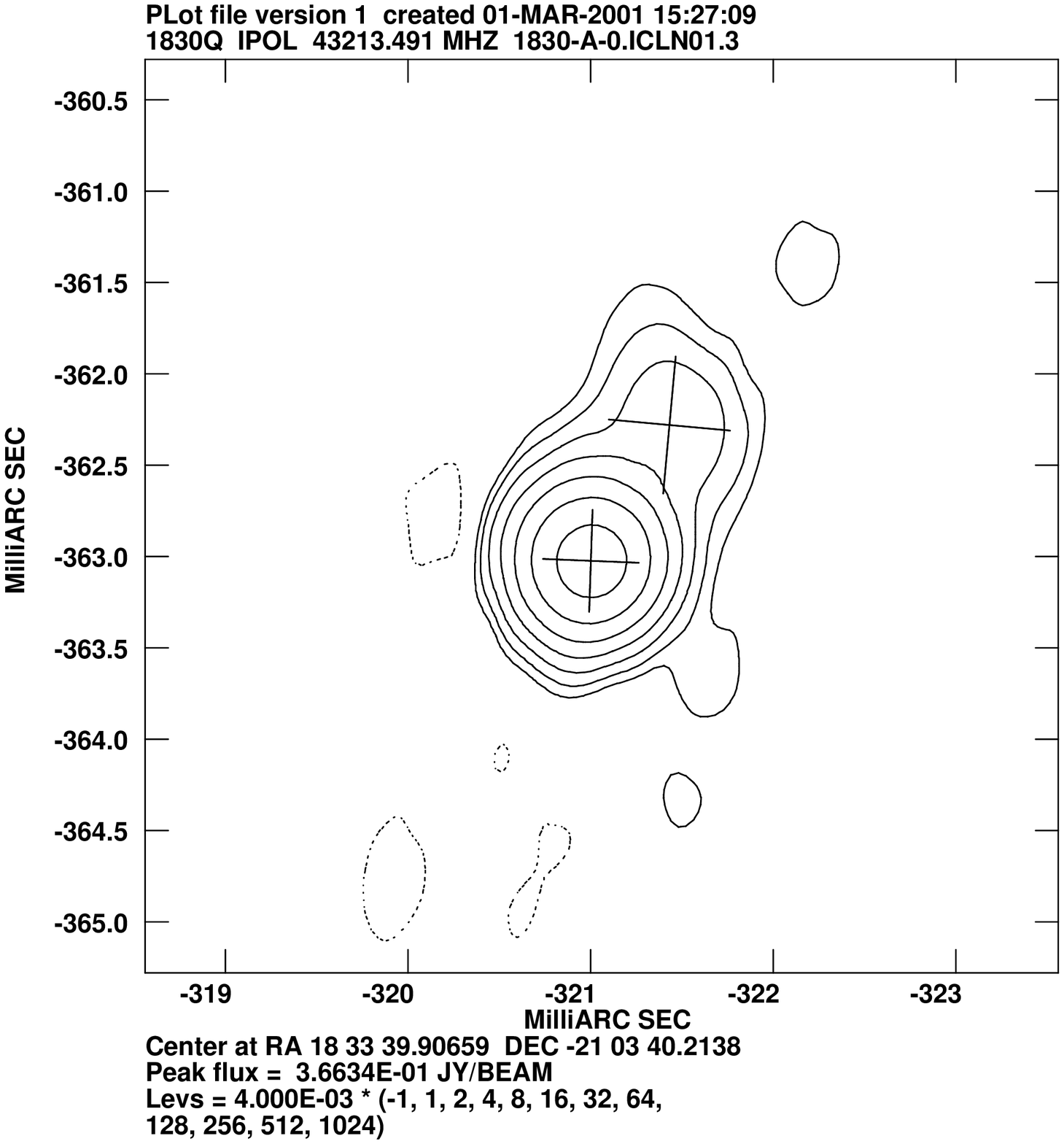}}
\put(-130,390){\includegraphics{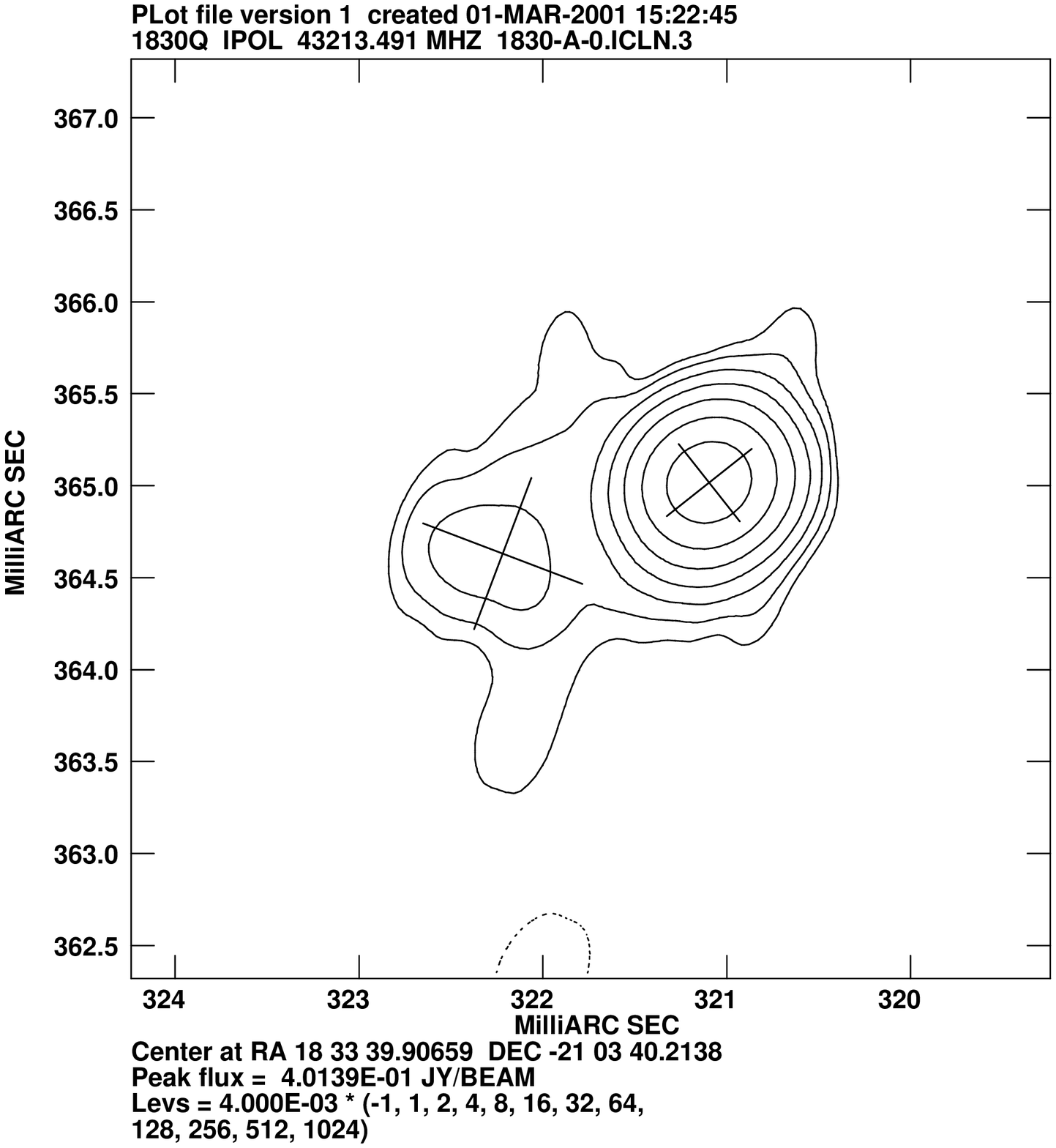}}

\put(-200,530){\includegraphics{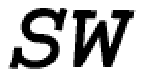}}
\put(-90,530){\includegraphics{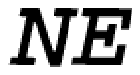}}

\put(-245,500){\includegraphics{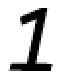}}
\put(-245,370){\includegraphics{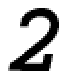}}
\put(-245,240){\includegraphics{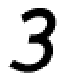}}
\put(-245,110){\includegraphics{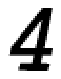}}

\put(25,500){\includegraphics{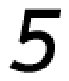}}
\put(25,370){\includegraphics{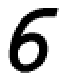}}
\put(25,240){\includegraphics{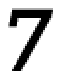}}
\put(25,110){\includegraphics{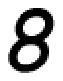}}

\put(70,530){\includegraphics{./sw.ps}}
\put(180,530){\includegraphics{./ne.ps}}

\put(30,0){\includegraphics{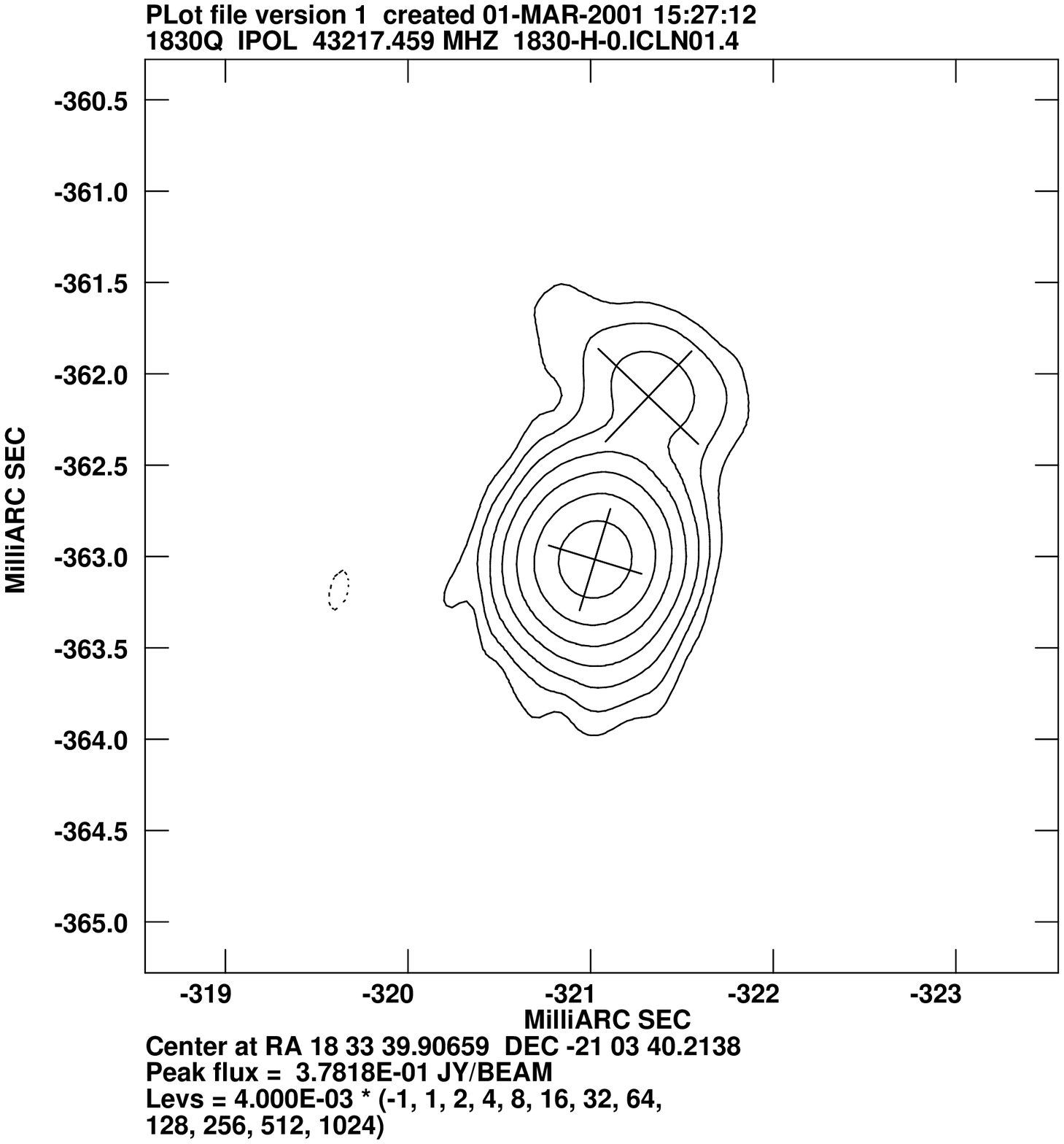}}
\put(140,0){\includegraphics{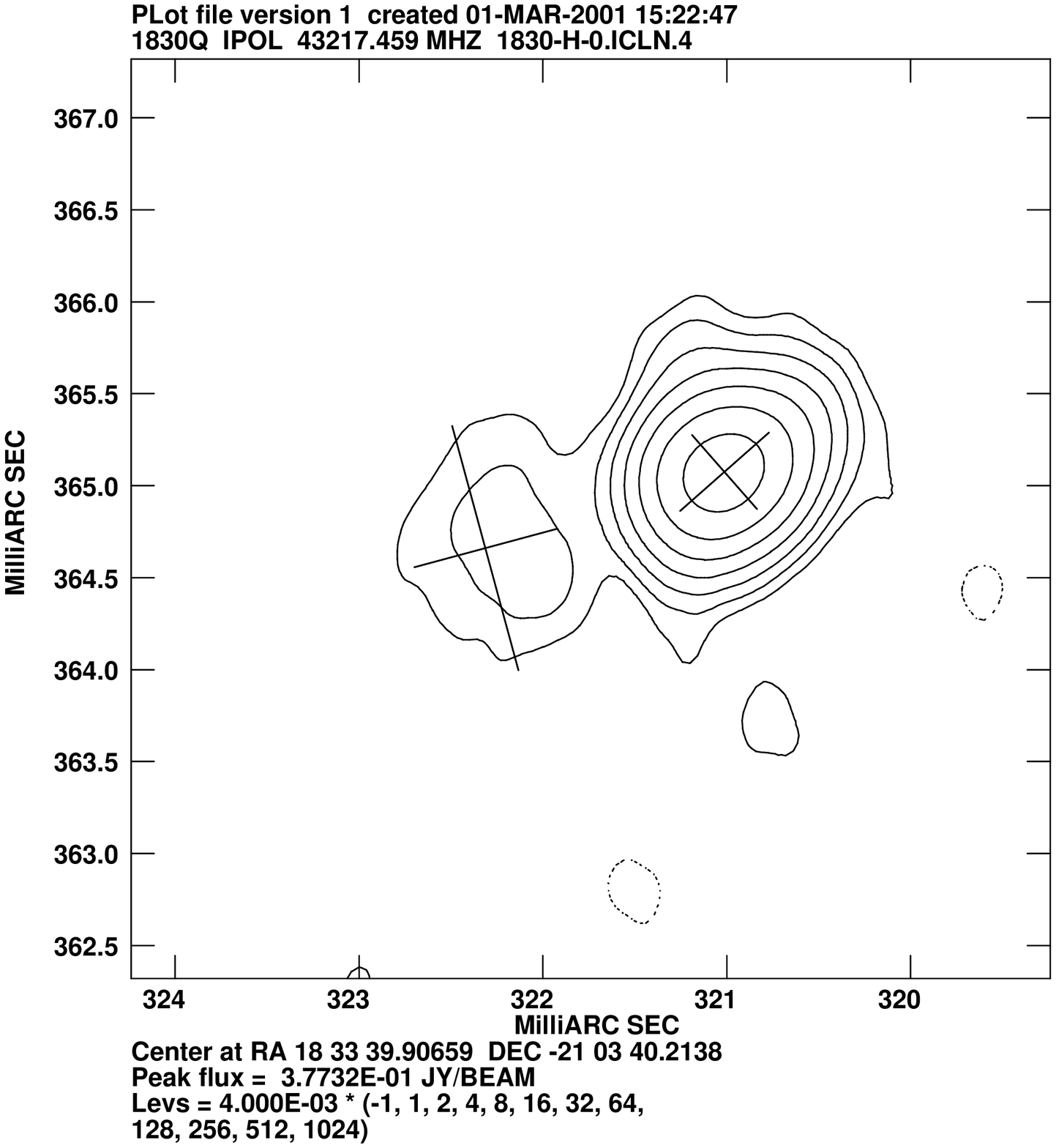}}
\put(30,130){\includegraphics{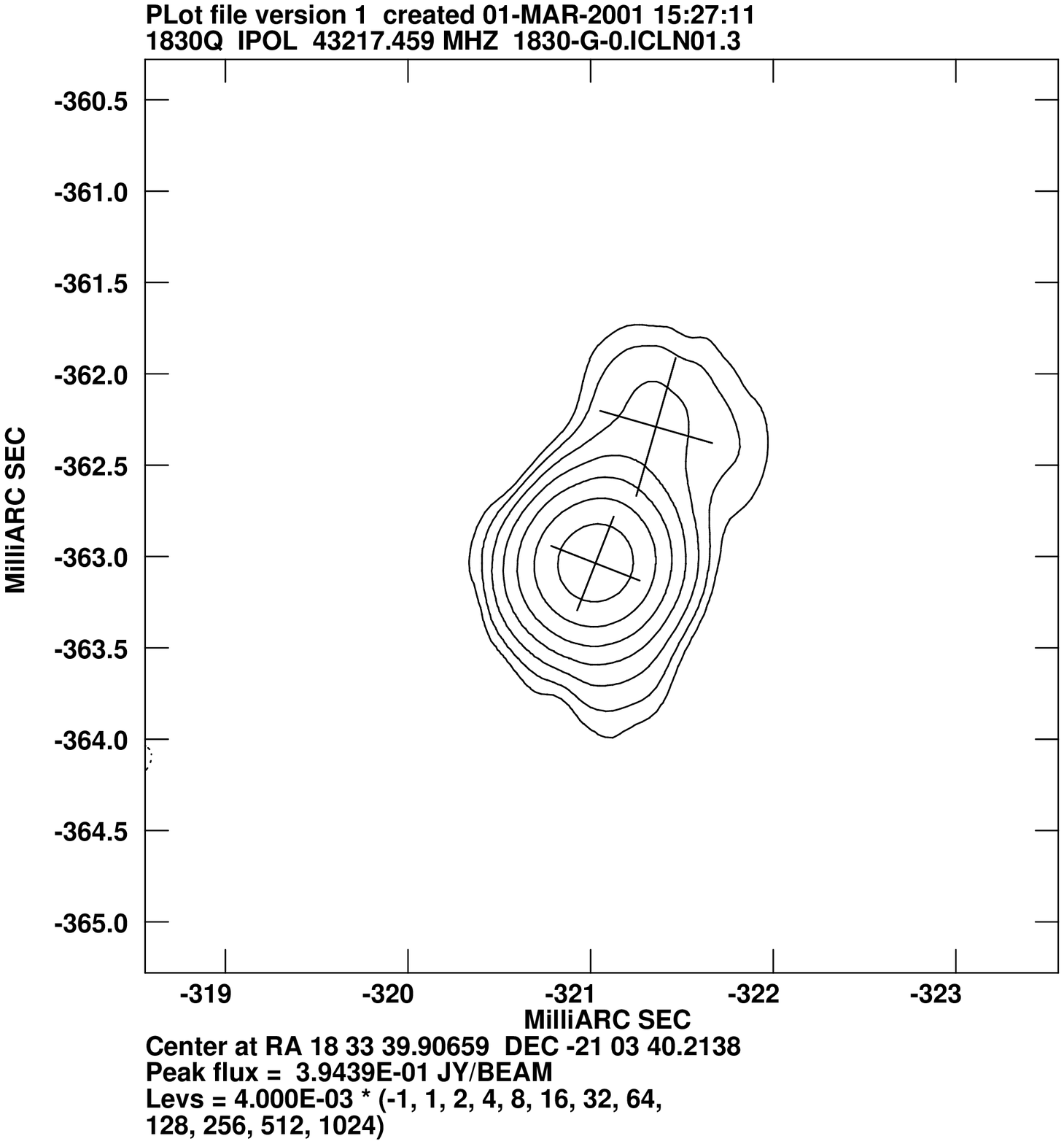}}
\put(140,130){\includegraphics{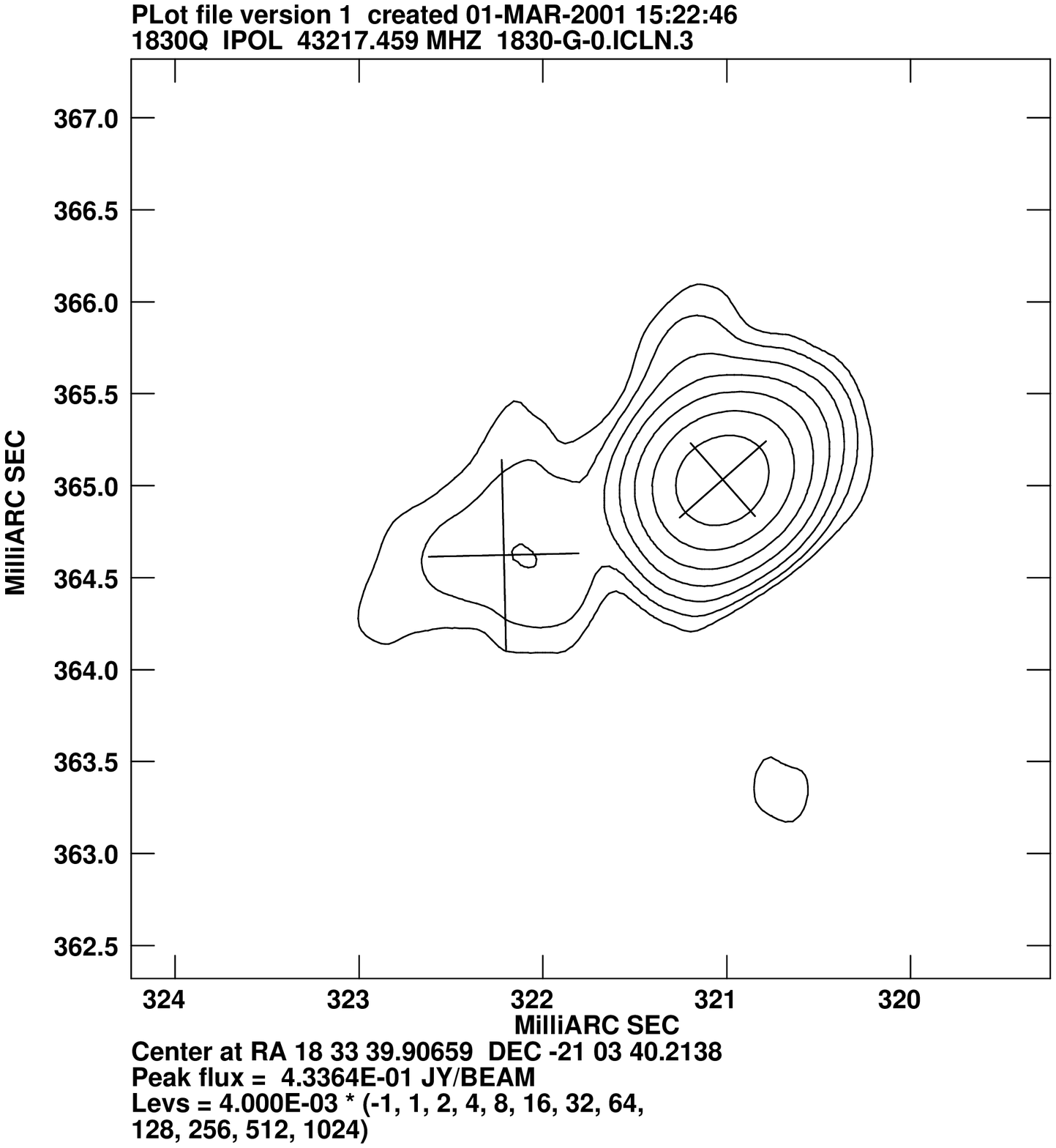}}
\put(30,260){\includegraphics{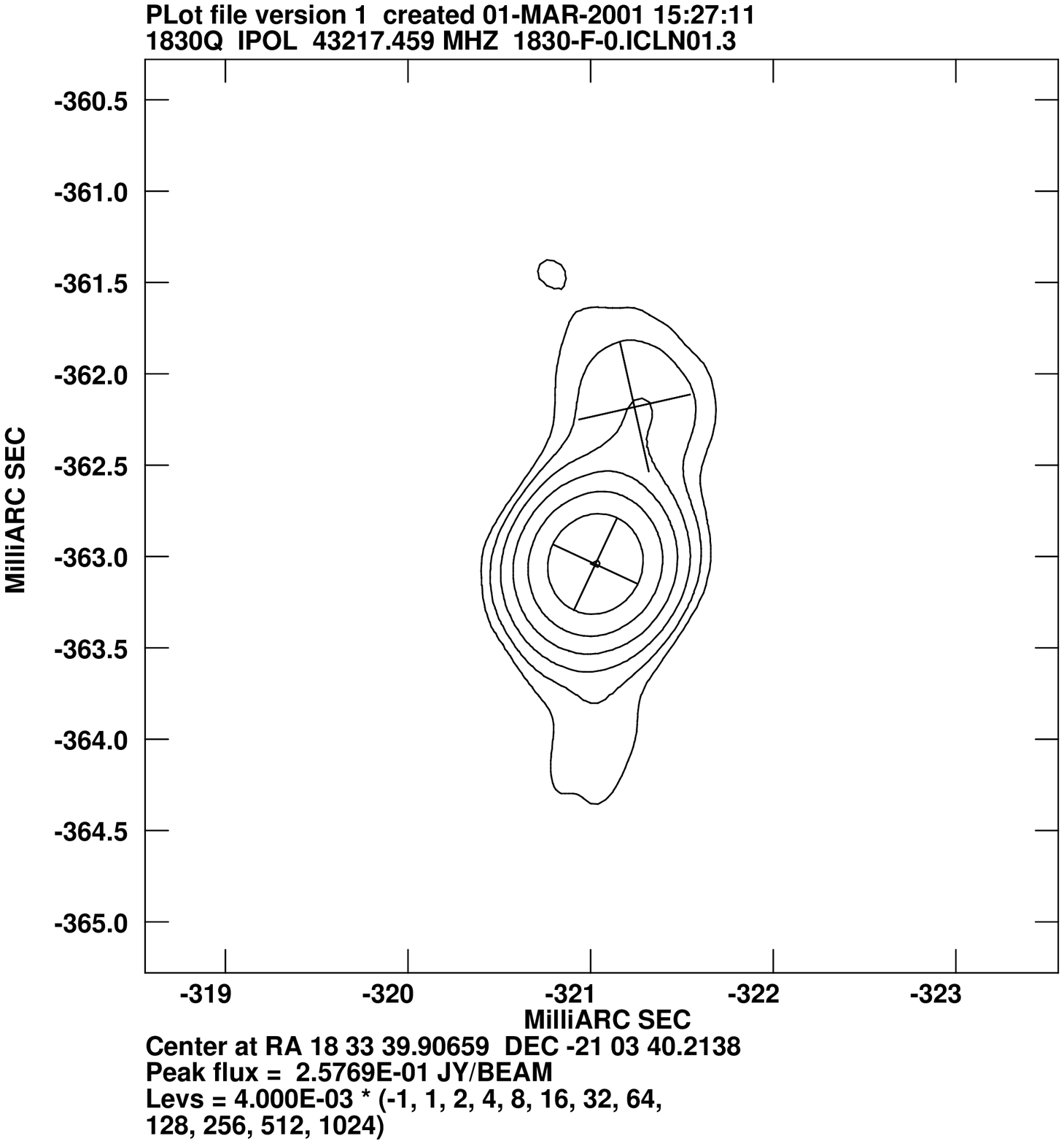}}
\put(140,260){\includegraphics{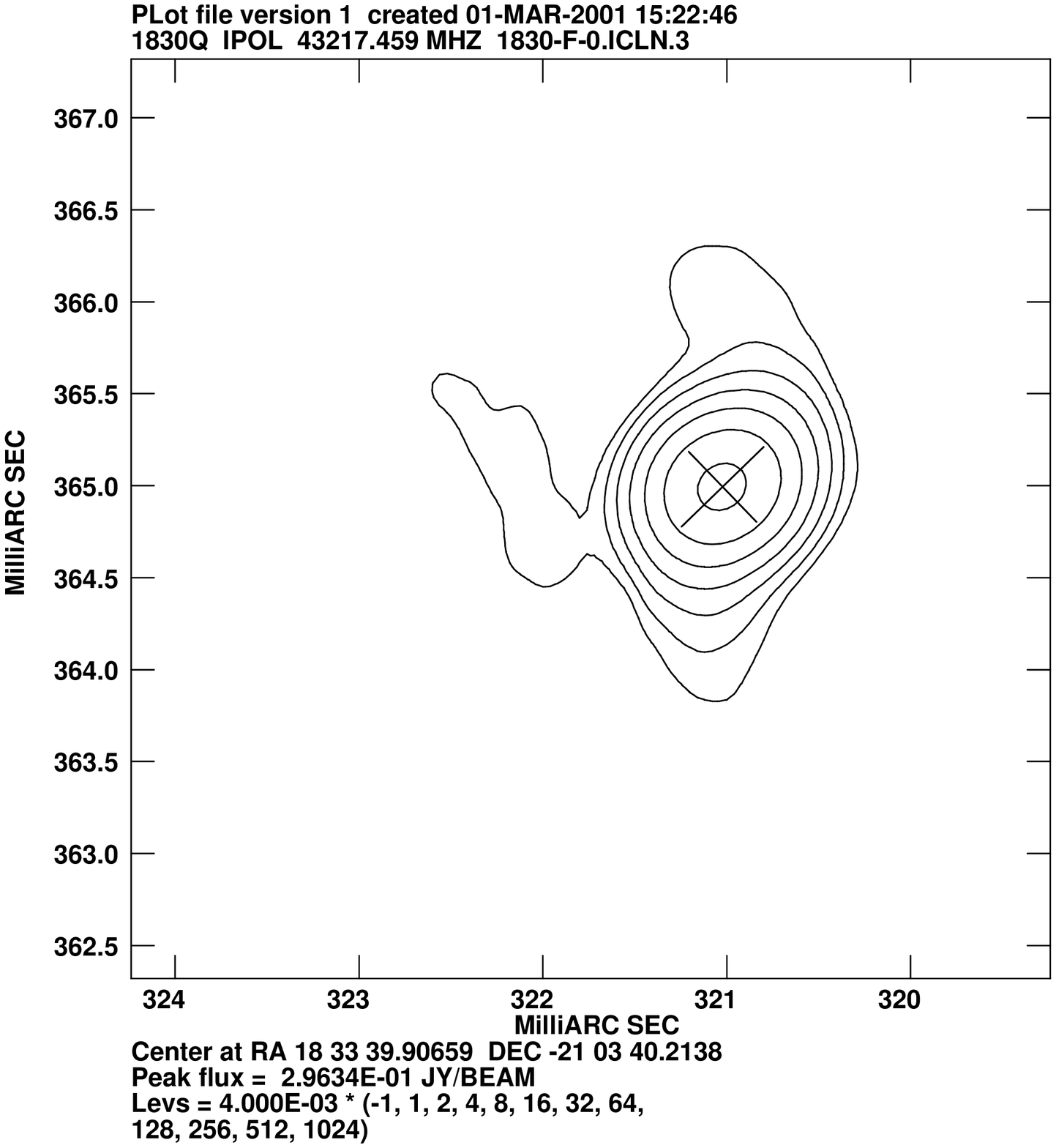}}
\put(30,390){\includegraphics{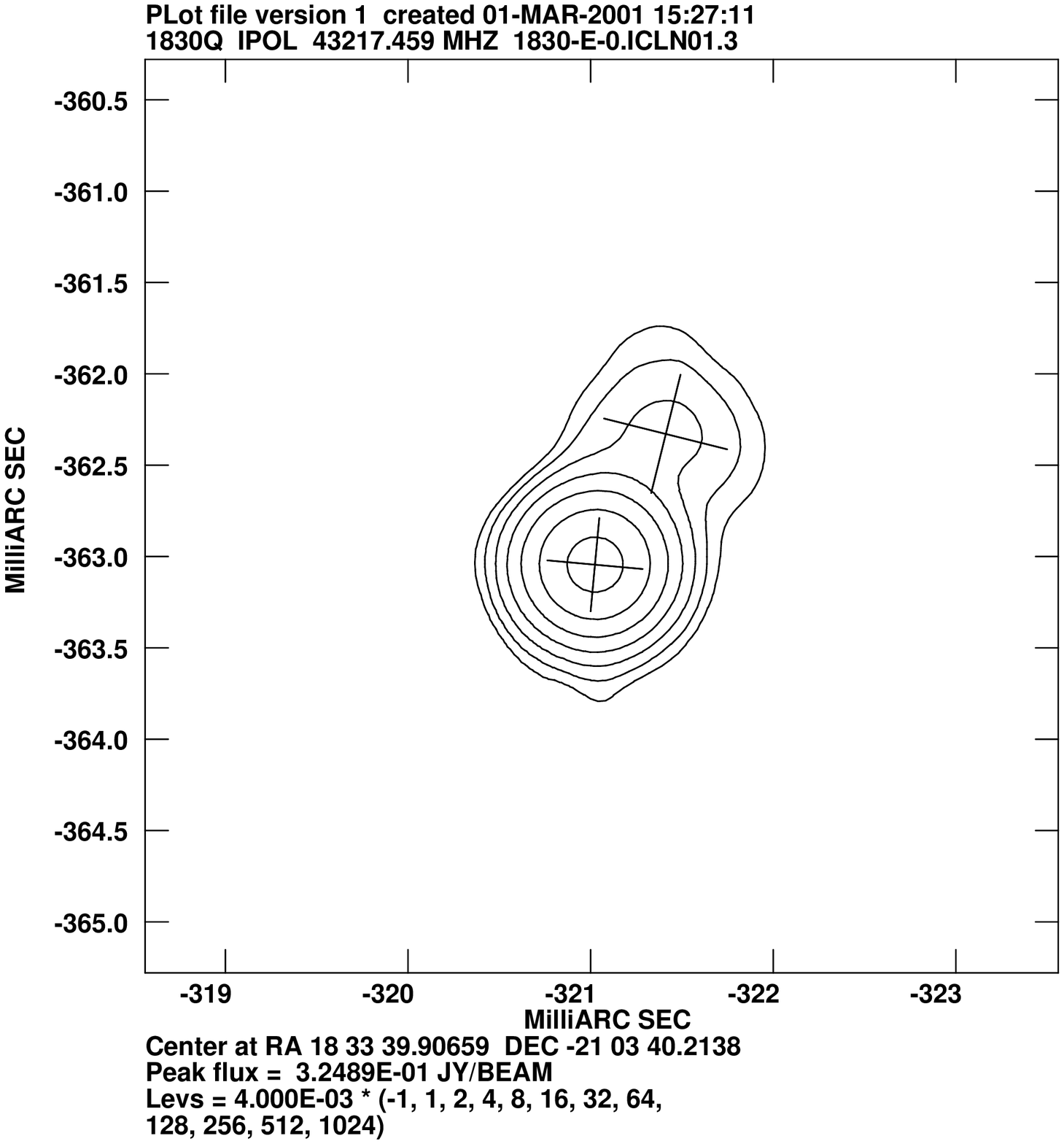}}
\put(140,390){\includegraphics{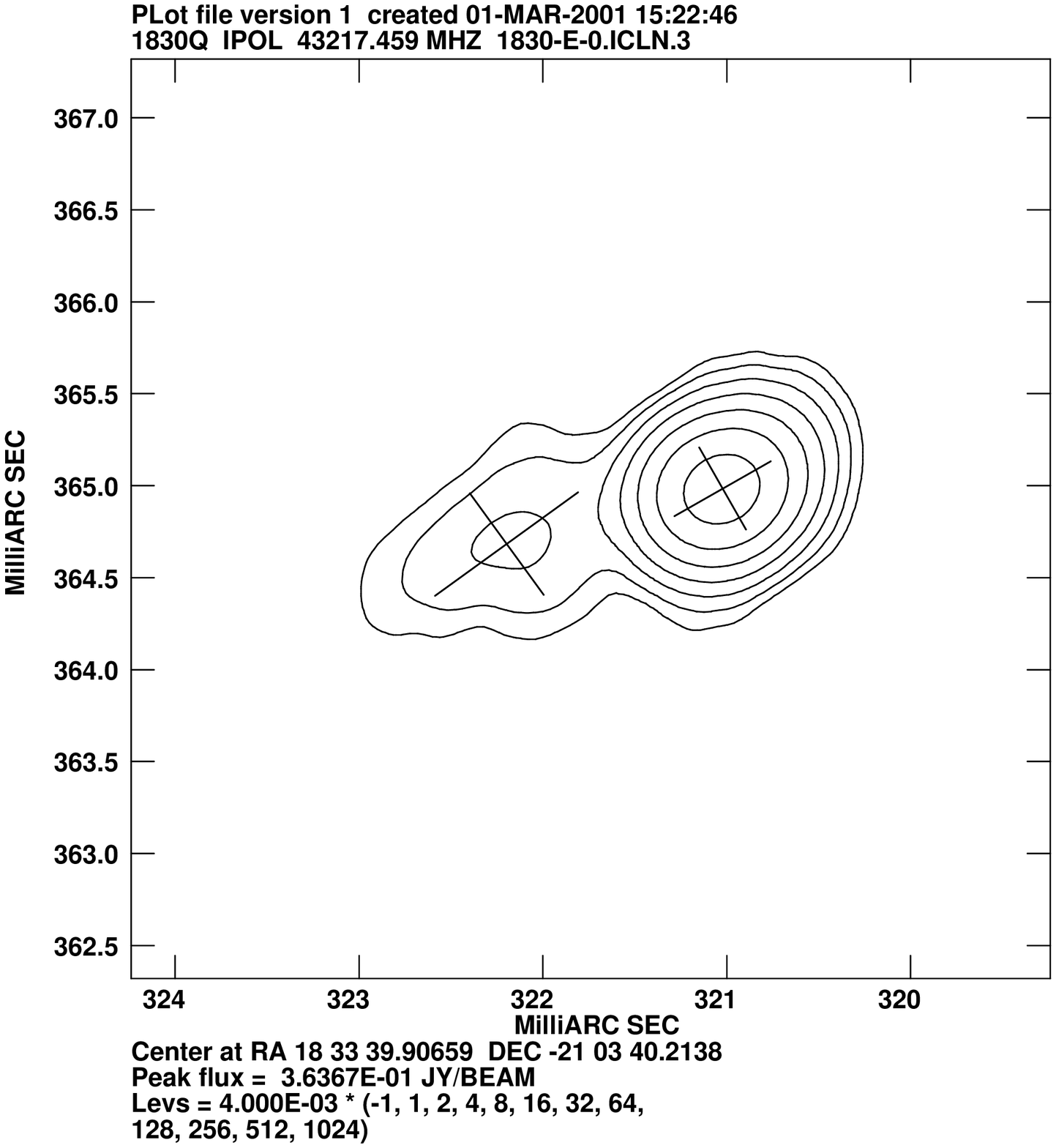}}
\end{picture}
\caption{Total intensity contour maps of the SW and NE images for all
eight epochs. The epoch of observation increases from top to bottom,
left to right. Contours are spaced by factors of two in brightness, with
the lowest at five times the r.m.s. noise of 0.8 mJy per beam. All
maps are on the same scale. The FWHM of the circular restoring beam is
0.5 mas. The crosses superimposed on the maps represent the position
and extent of the fitted Gaussian components. The measured time-delay of
$26^{+4}_{-5}$ days implies that structure in the NE image at one
epoch corresponds to that in the SW image roughly two epochs later. }
\label{fig1}
\end{figure*}

\begin{figure*}
\vspace{19cm}
\begin{picture}(40,40)
\put(-240,0){\includegraphics{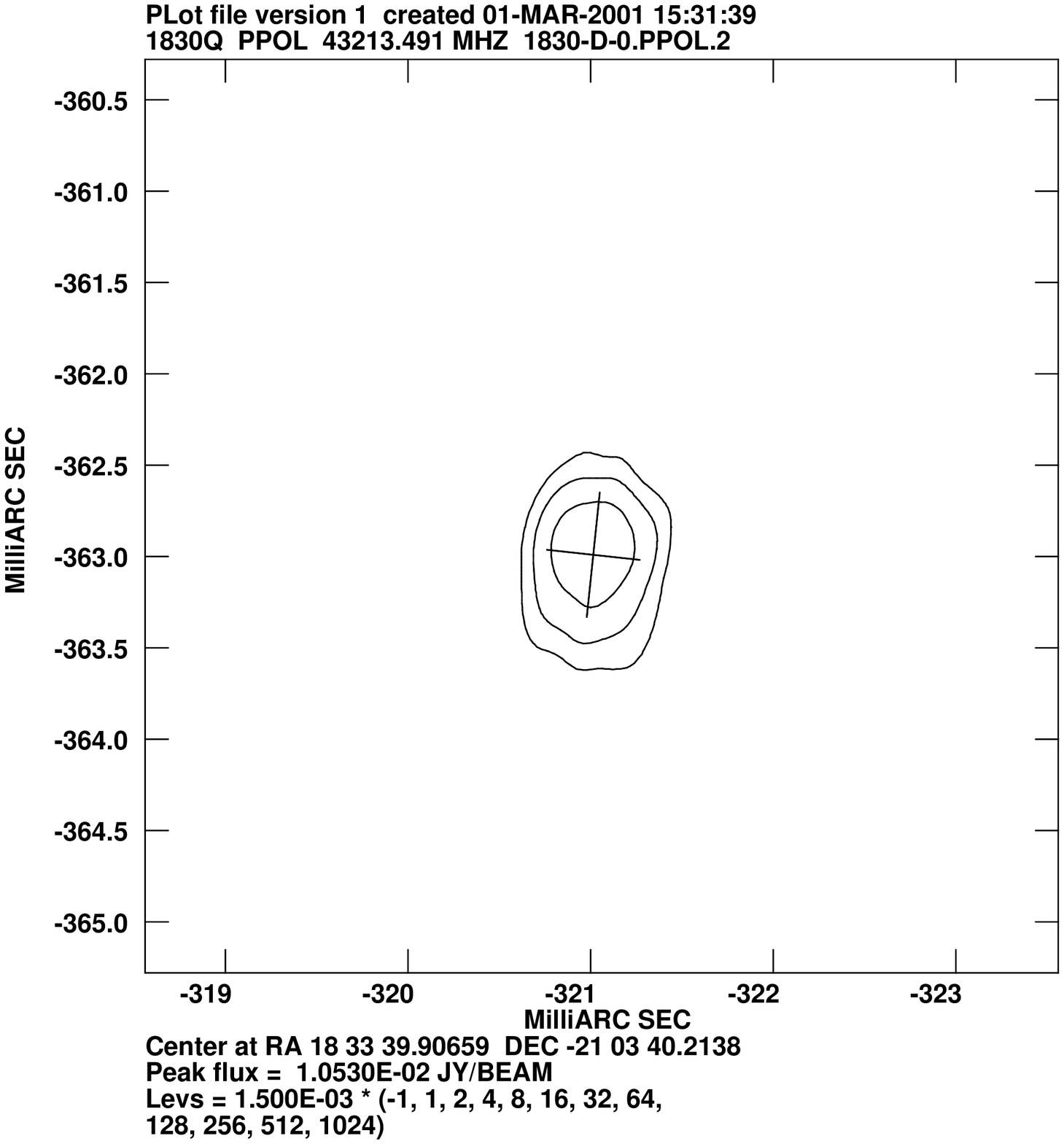}}
\put(-130,0){\includegraphics{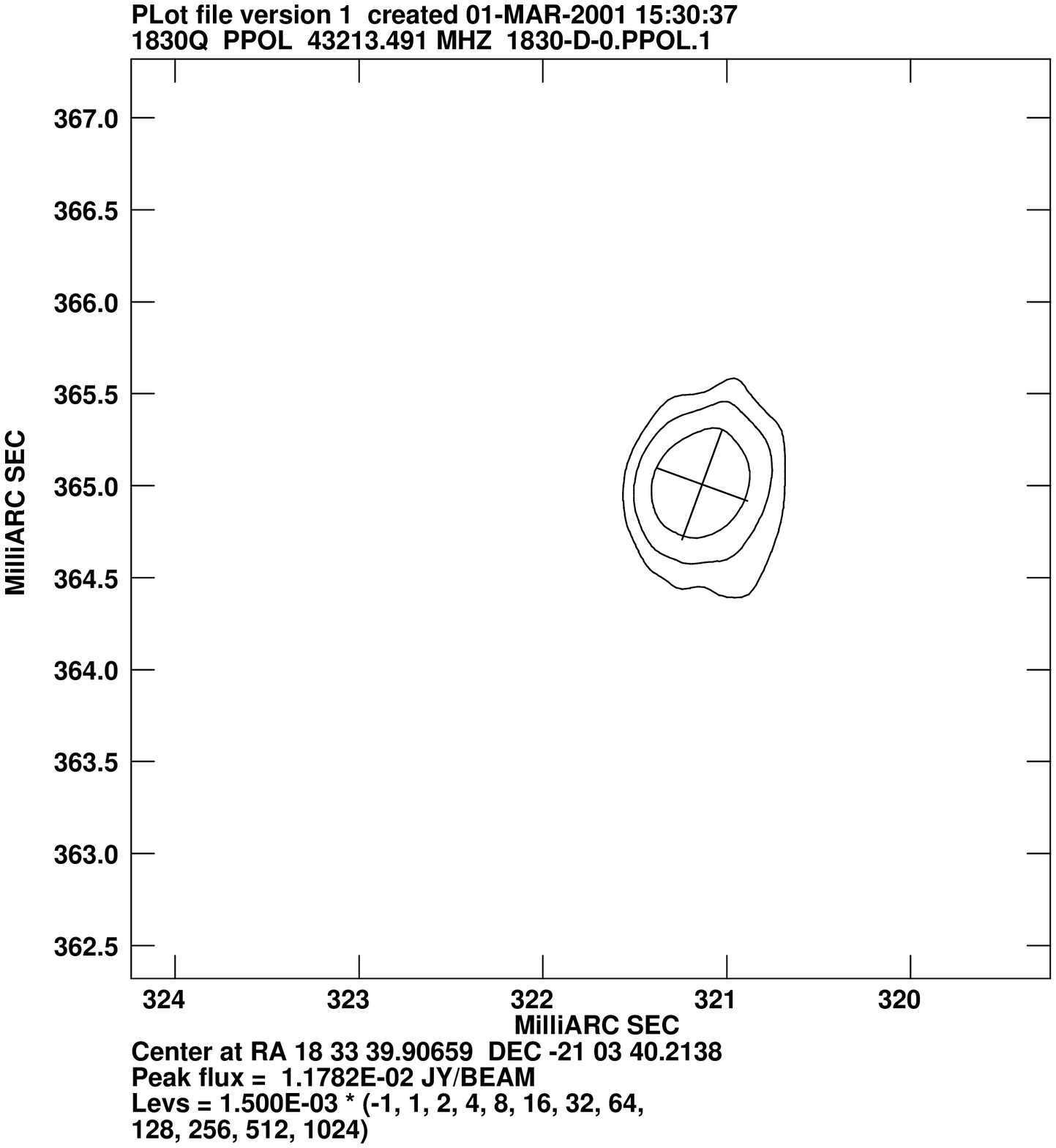}}
\put(-240,130){\includegraphics{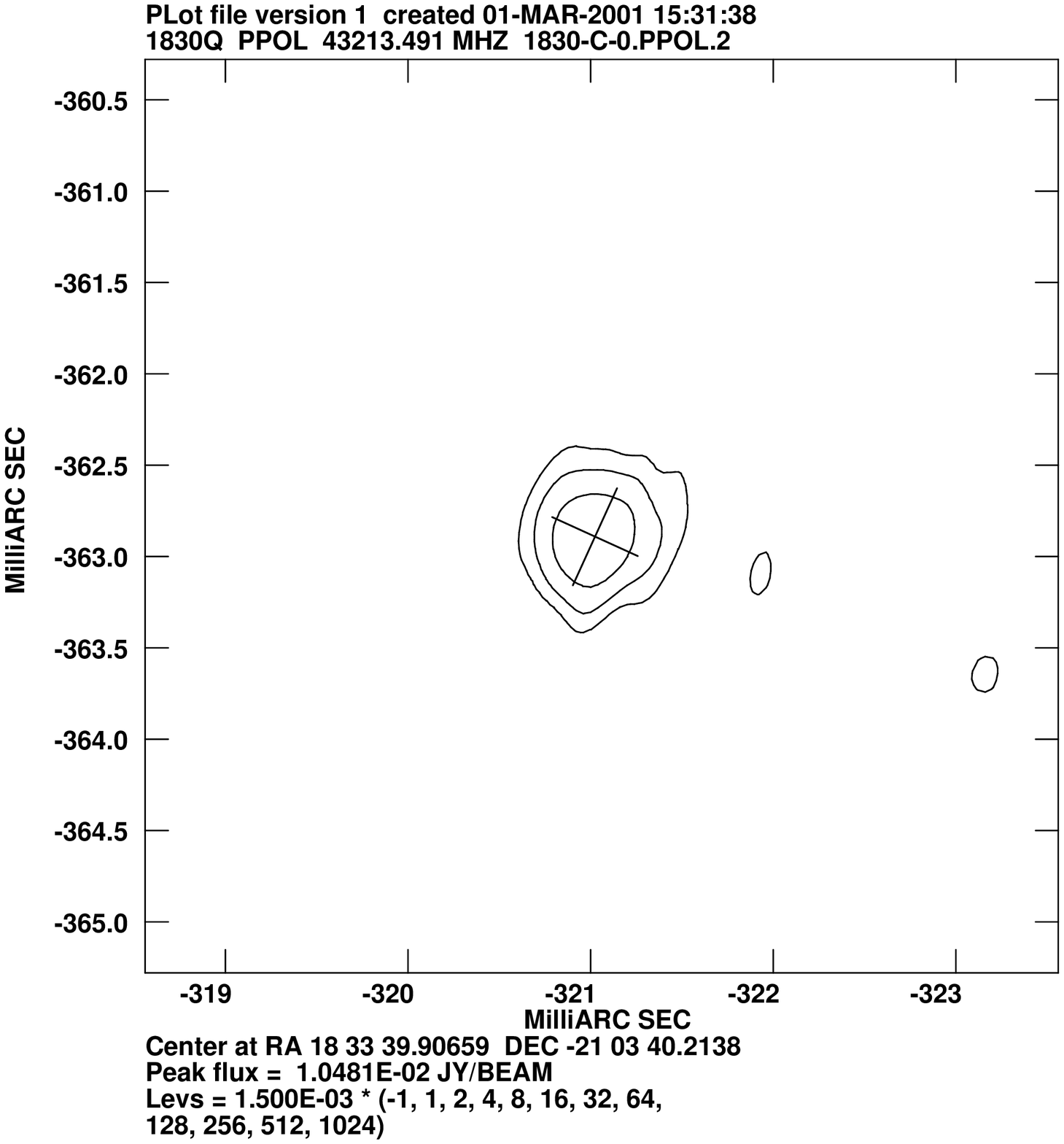}}
\put(-130,130){\includegraphics{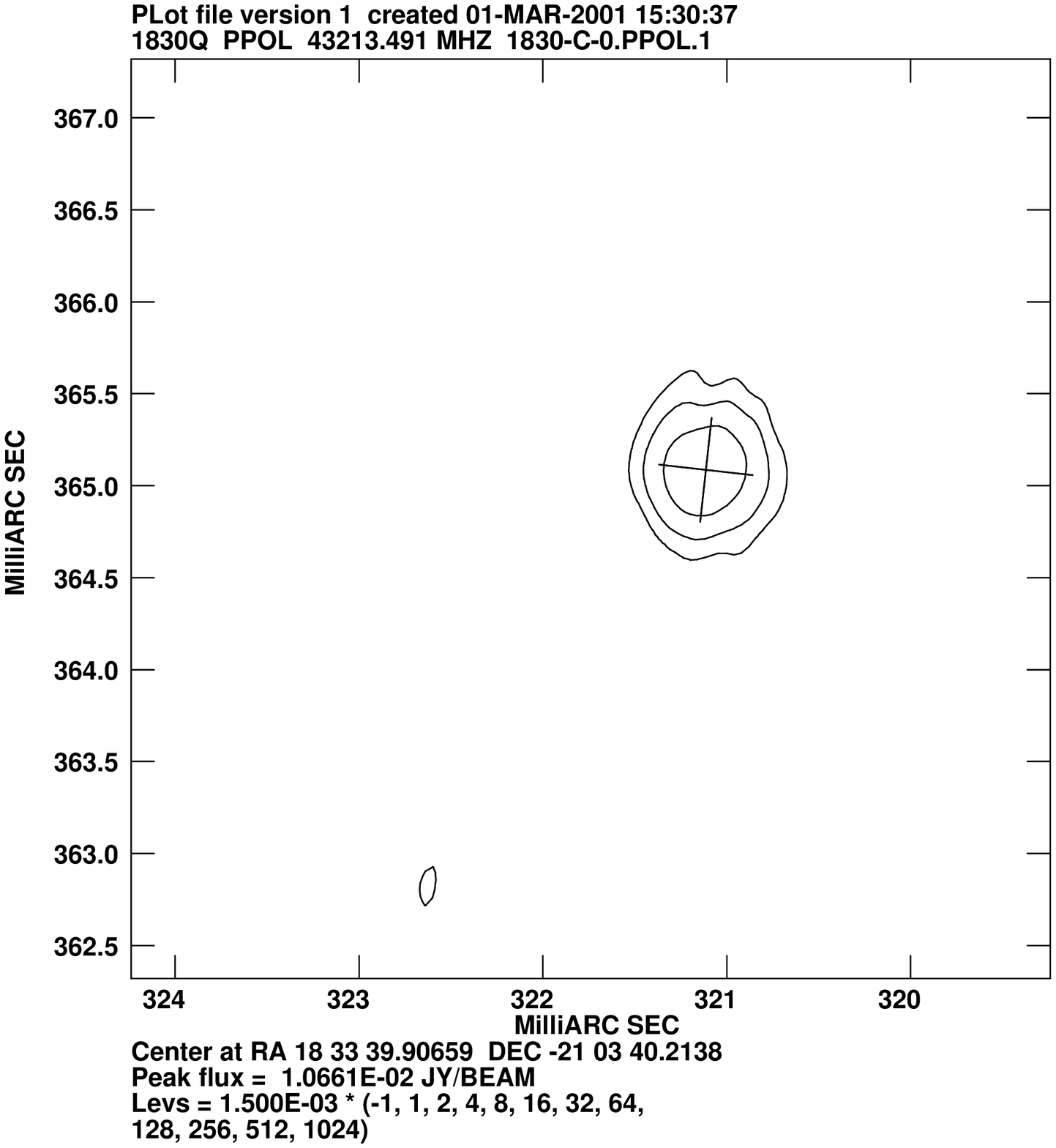}}
\put(-240,260){\includegraphics{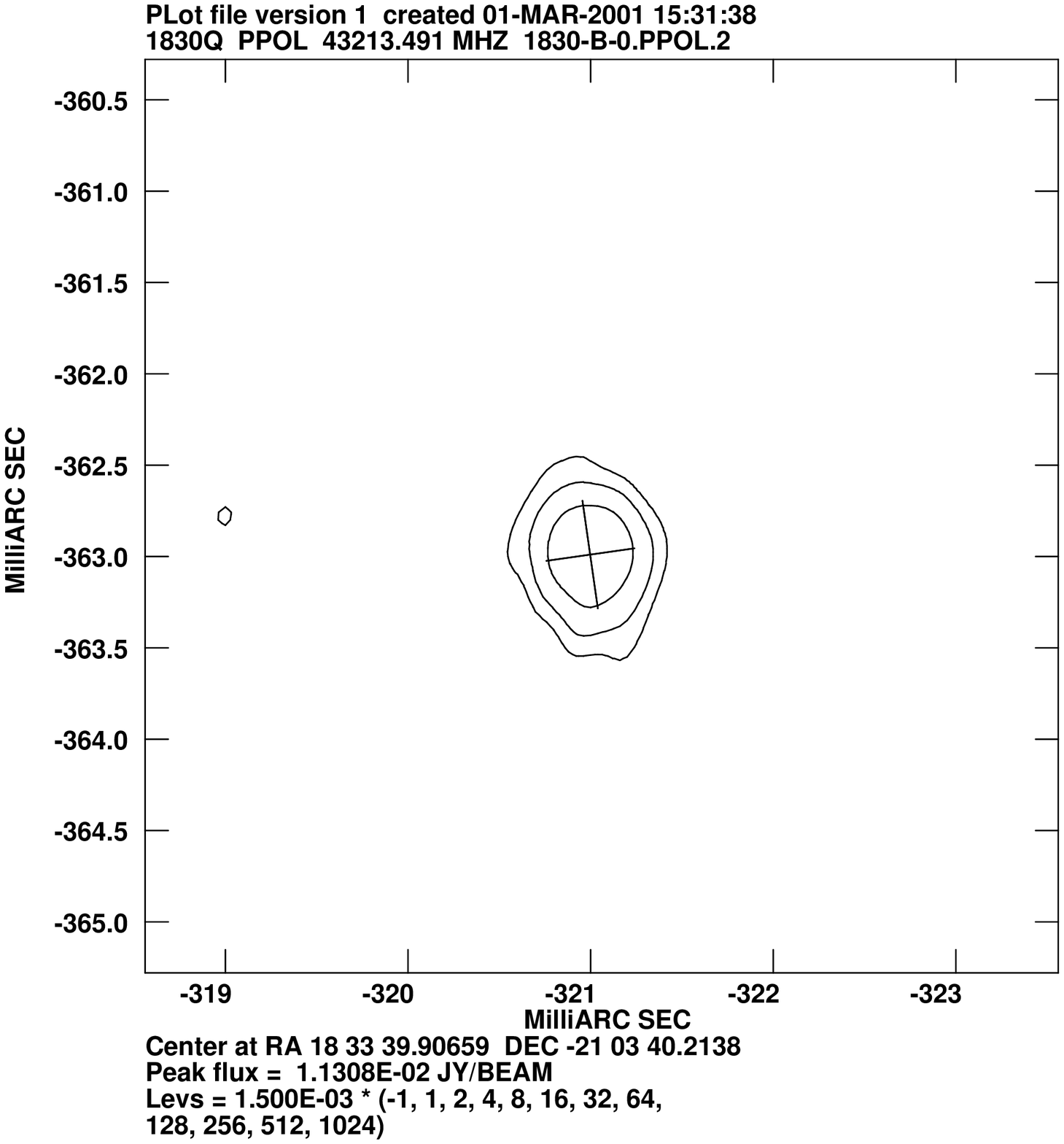}}
\put(-130,260){\includegraphics{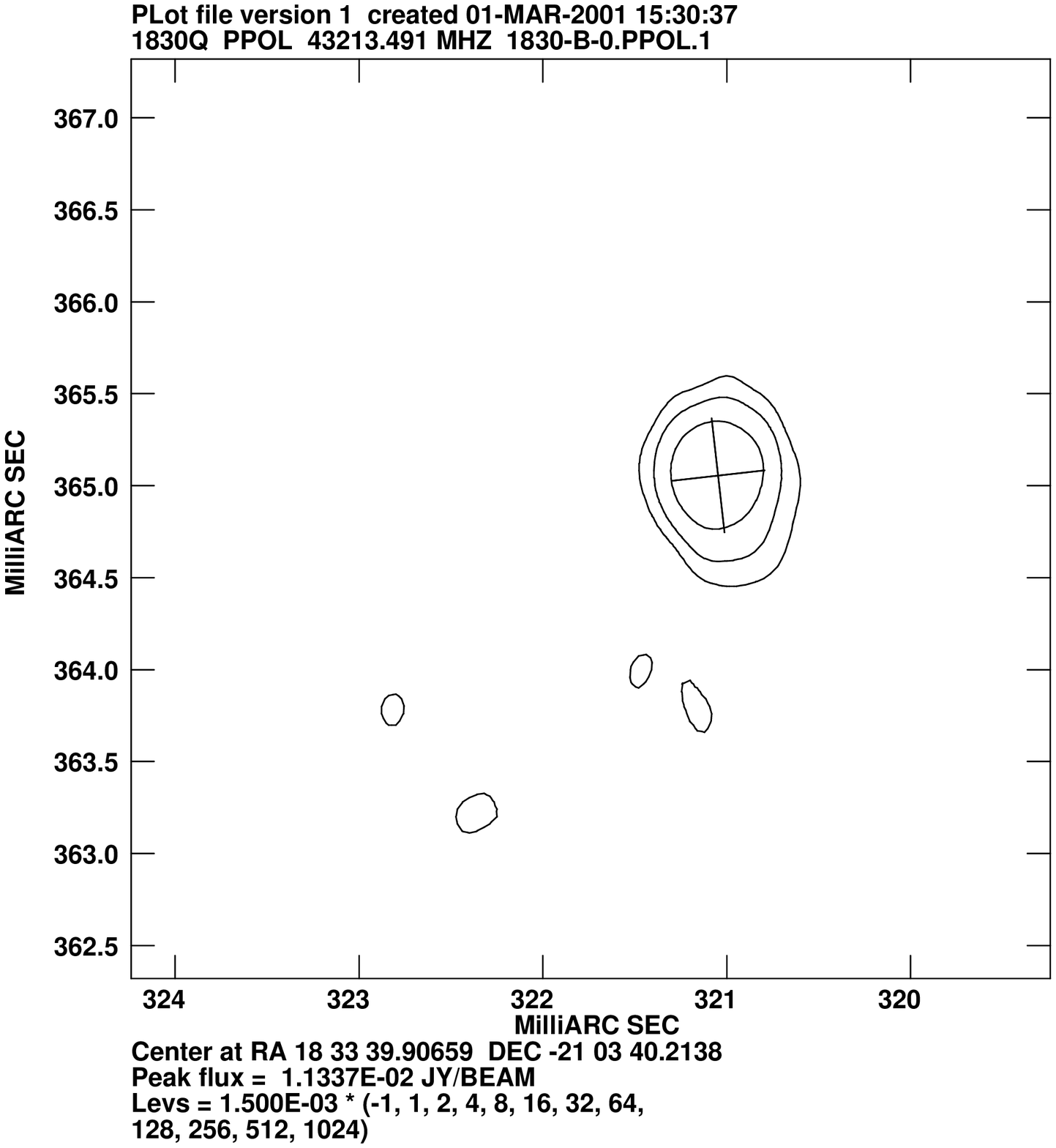}}
\put(-240,390){\includegraphics{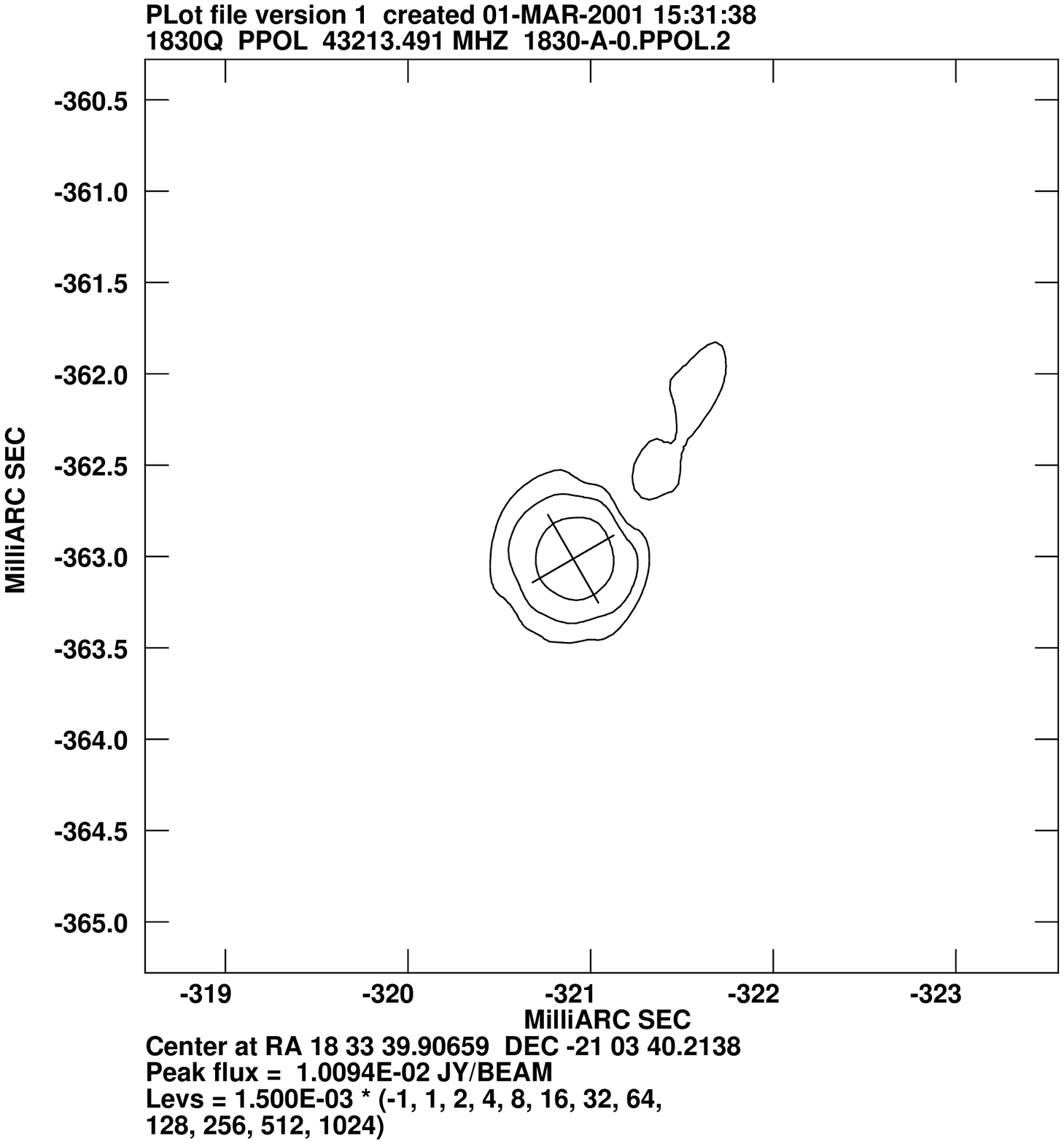}}
\put(-130,390){\includegraphics{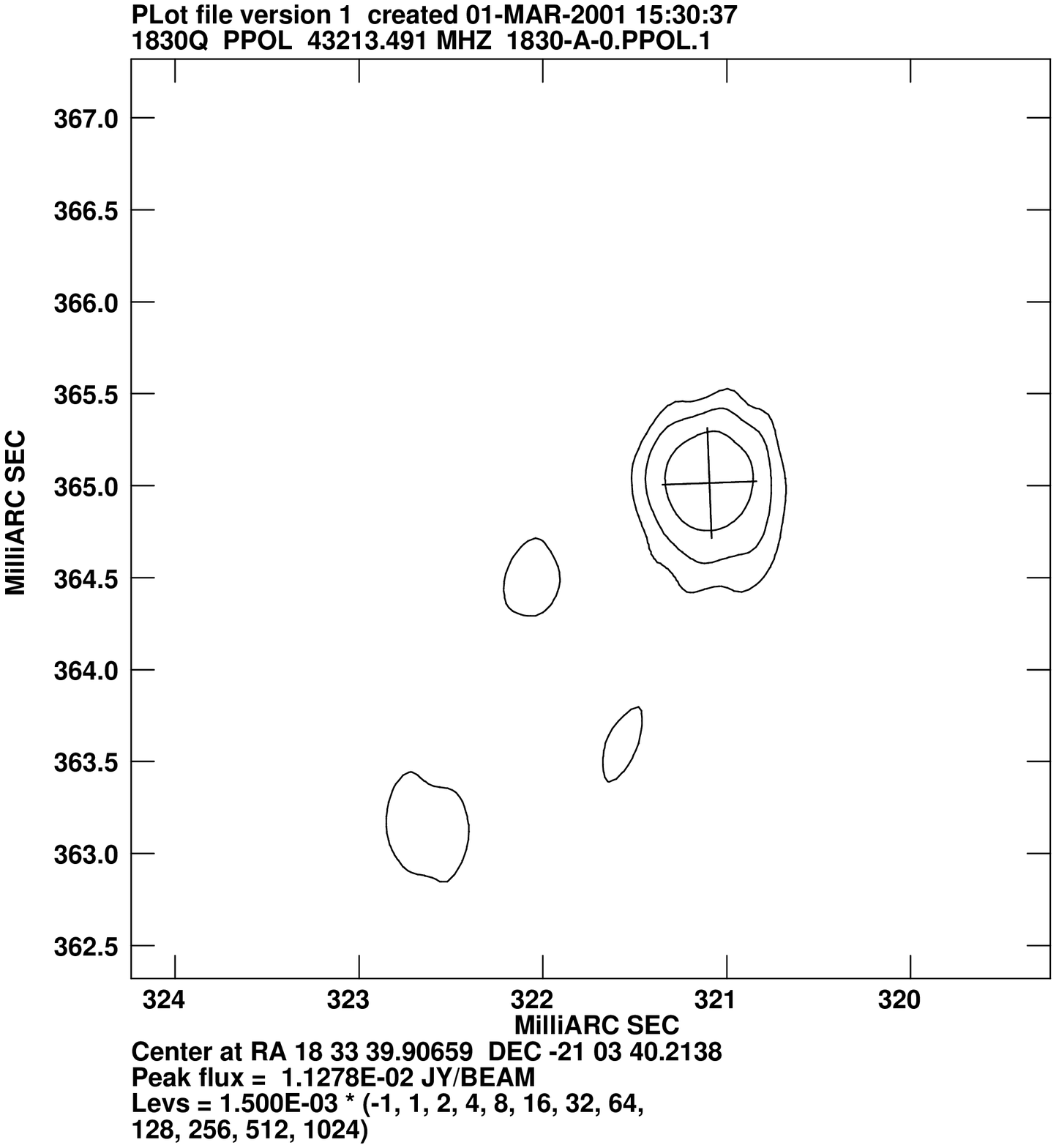}}

\put(-200,530){\includegraphics{./sw.ps}}
\put(-90,530){\includegraphics{./ne.ps}}

\put(-245,500){\includegraphics{./1.ps}}
\put(-245,370){\includegraphics{./2.ps}}
\put(-245,240){\includegraphics{./3.ps}}
\put(-245,110){\includegraphics{./4.ps}}

\put(25,500){\includegraphics{./5.ps}}
\put(25,370){\includegraphics{./6.ps}}
\put(25,240){\includegraphics{./7.ps}}
\put(25,110){\includegraphics{./8.ps}}

\put(70,530){\includegraphics{./sw.ps}}
\put(180,530){\includegraphics{./ne.ps}}

\put(30,0){\includegraphics{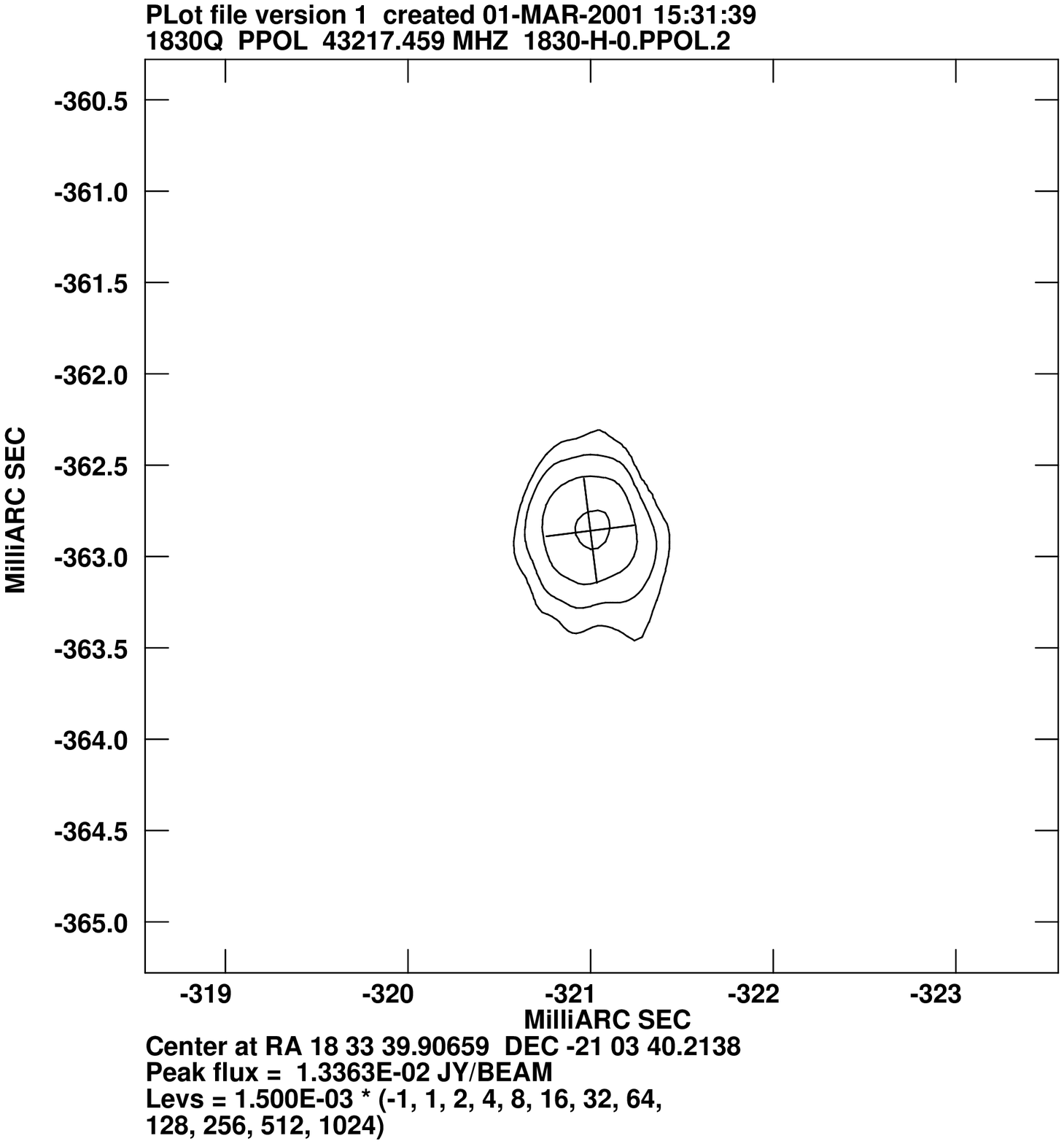}}
\put(140,0){\includegraphics{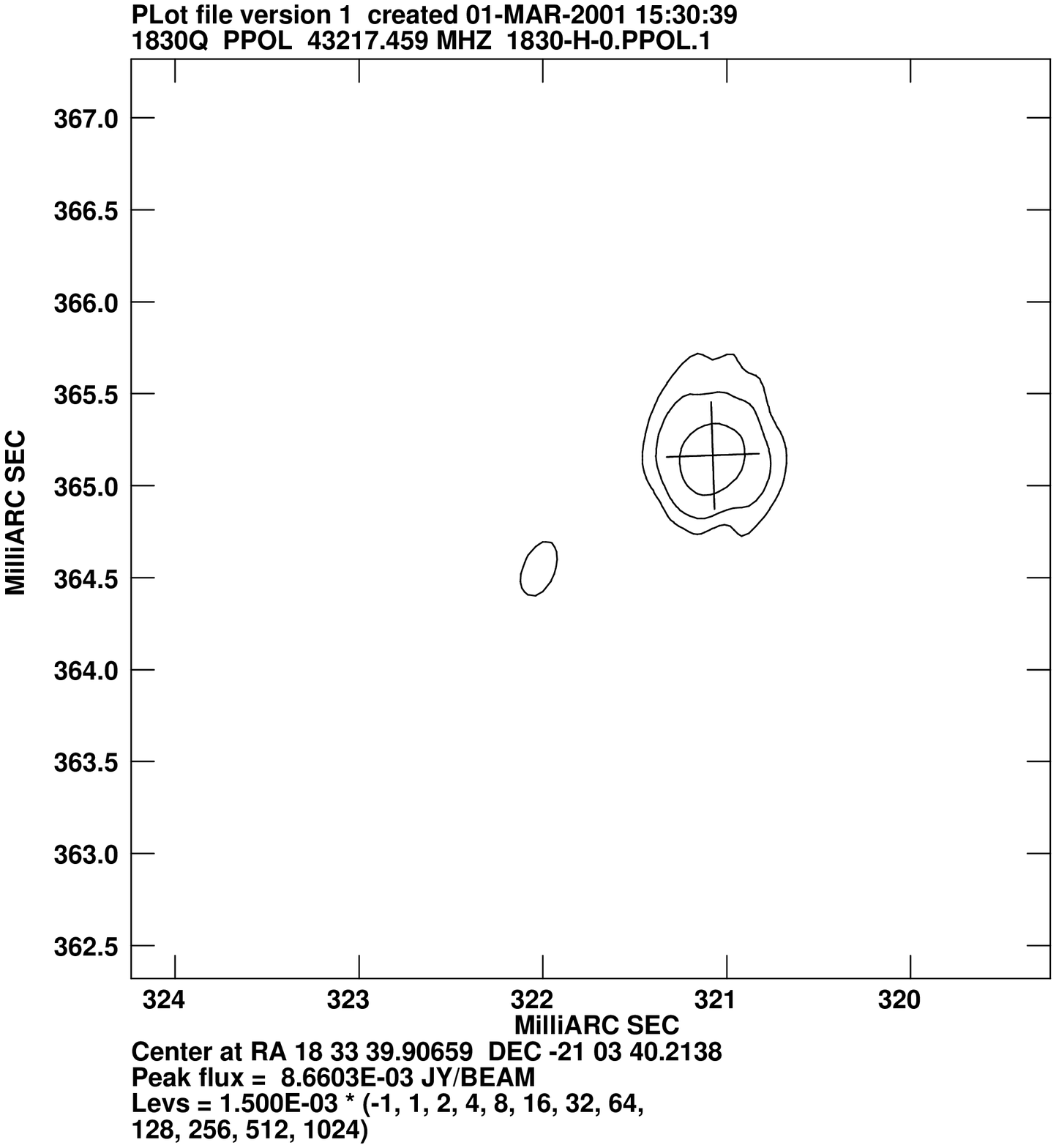}}
\put(30,130){\includegraphics{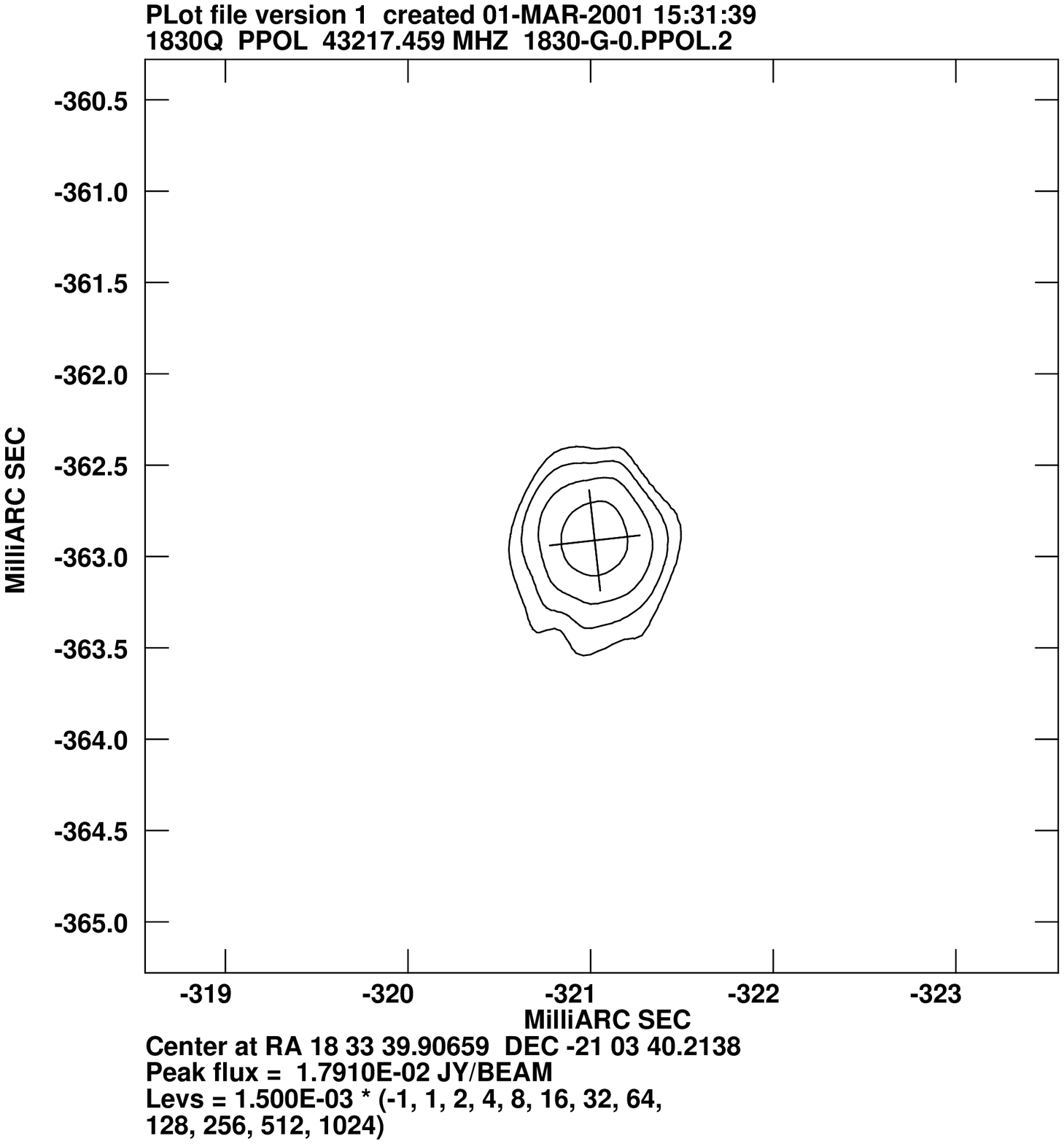}}
\put(140,130){\includegraphics{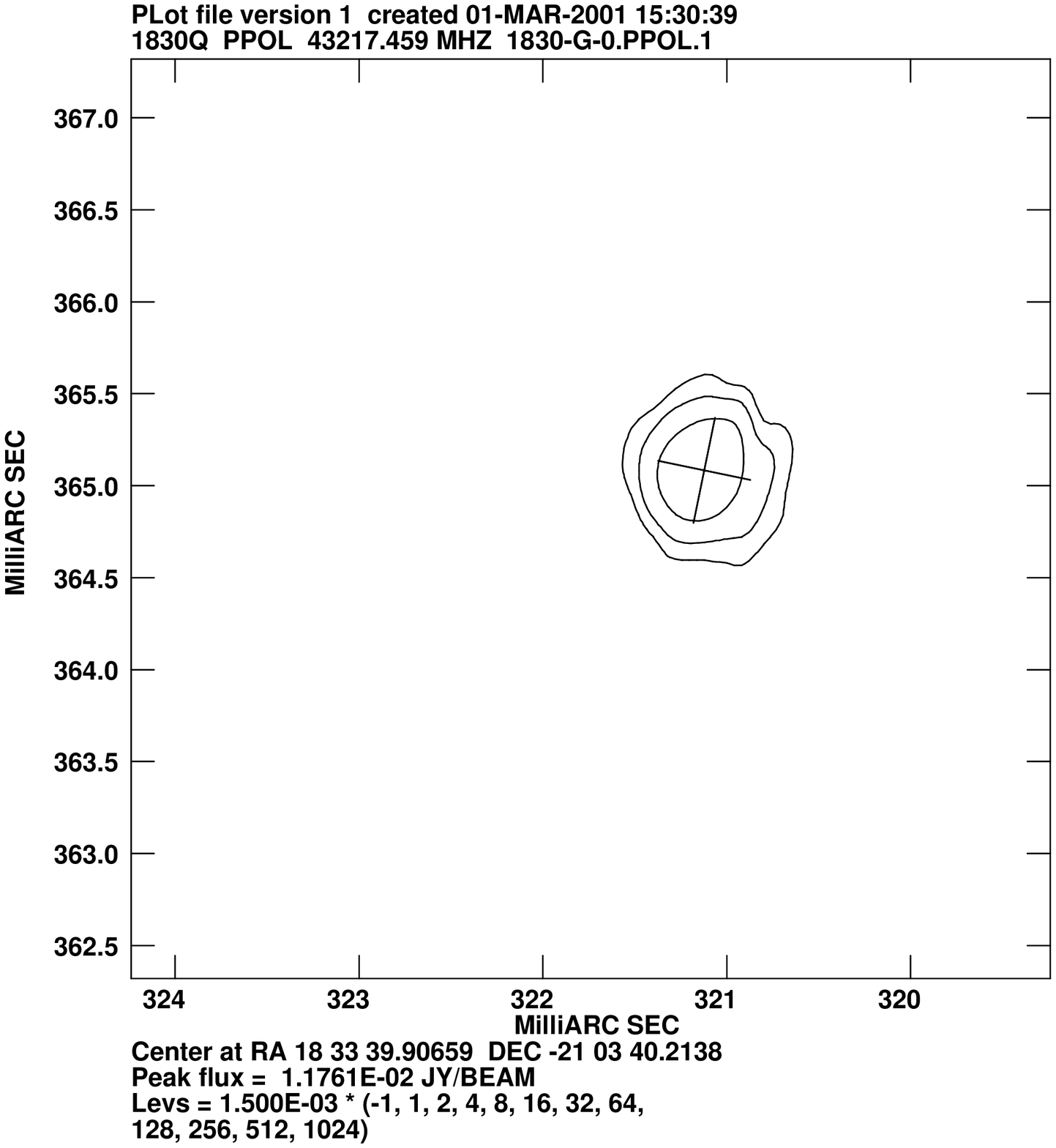}}
\put(30,260){\includegraphics{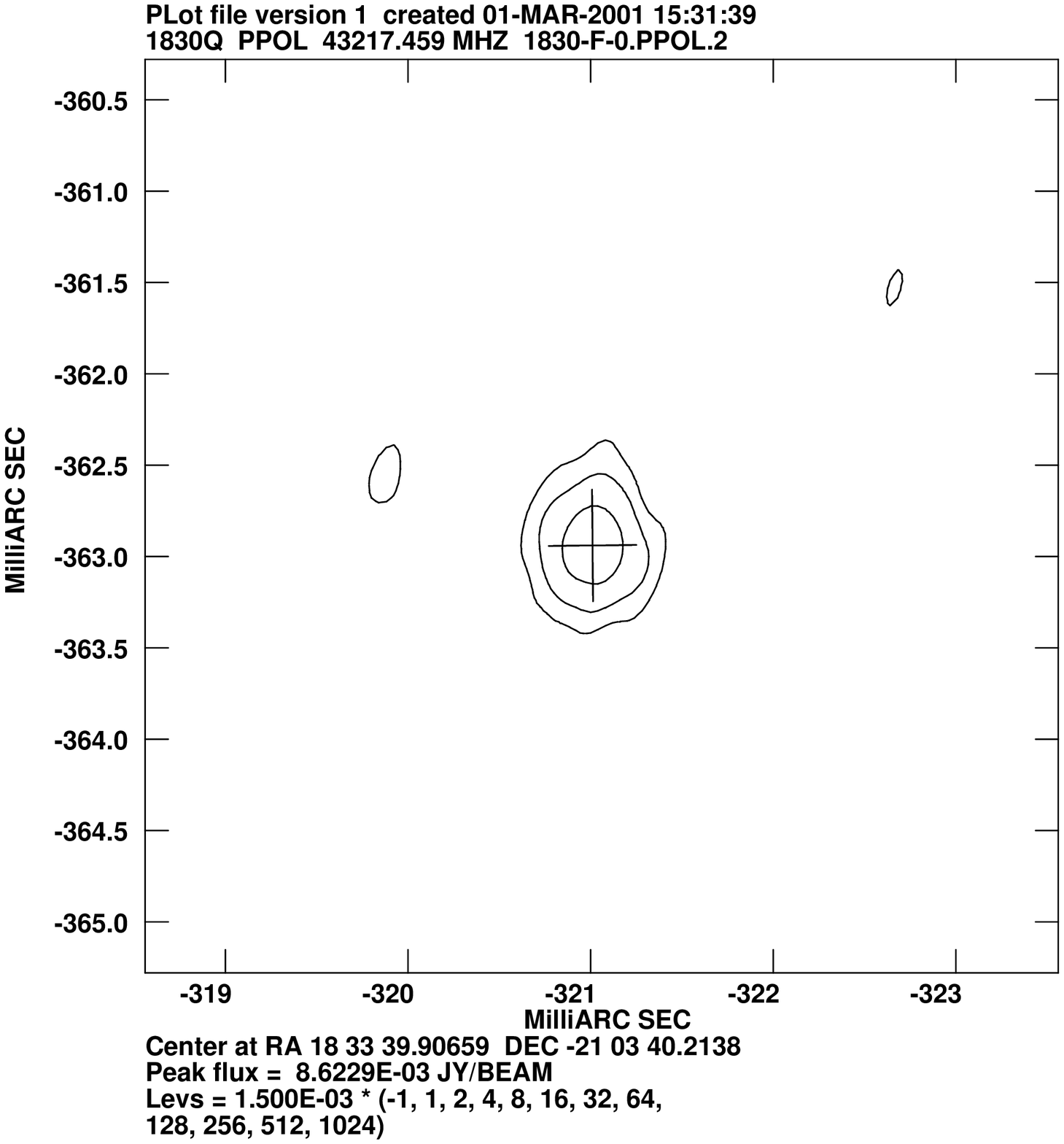}}
\put(140,260){\includegraphics{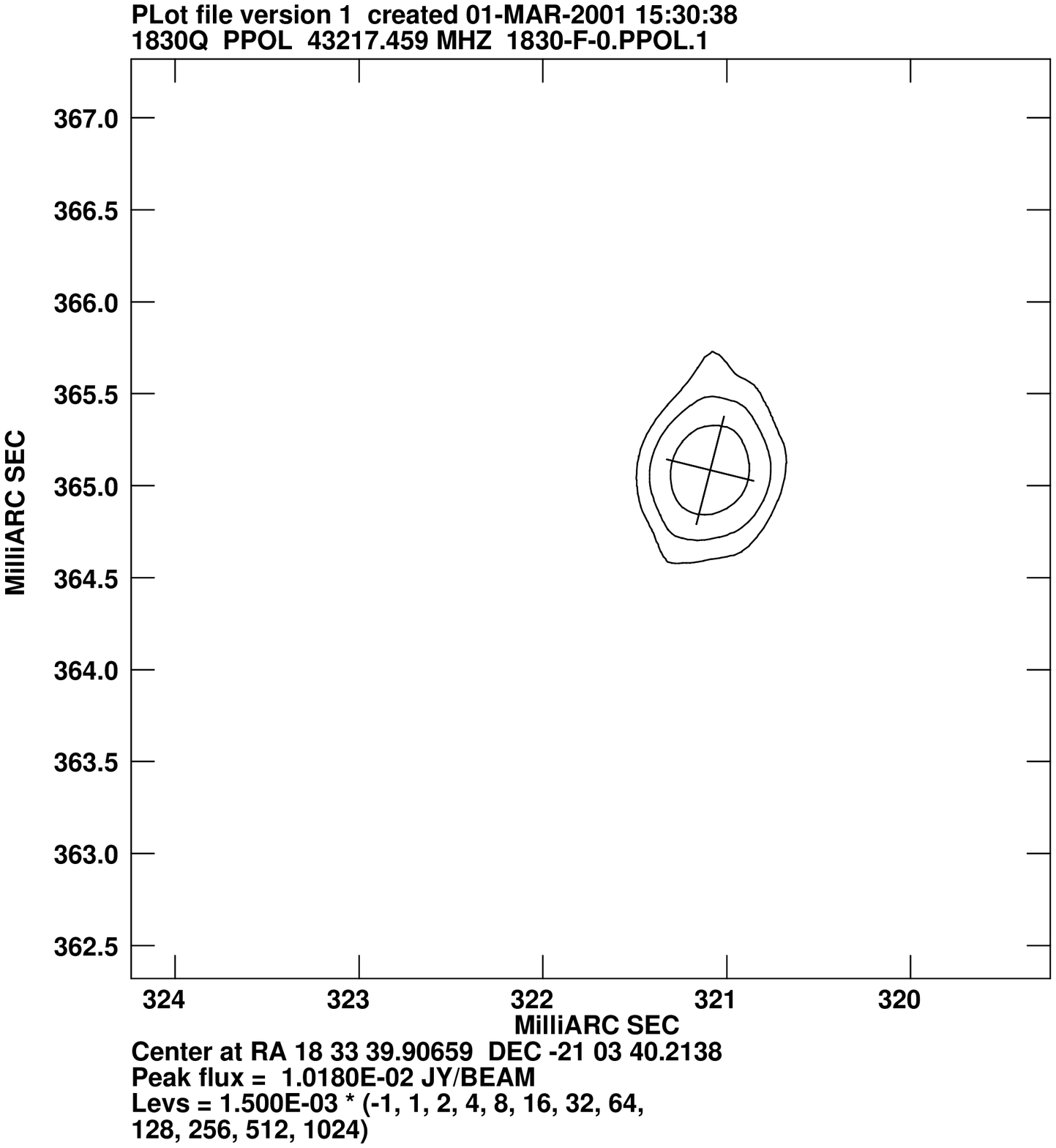}}
\put(30,390){\includegraphics{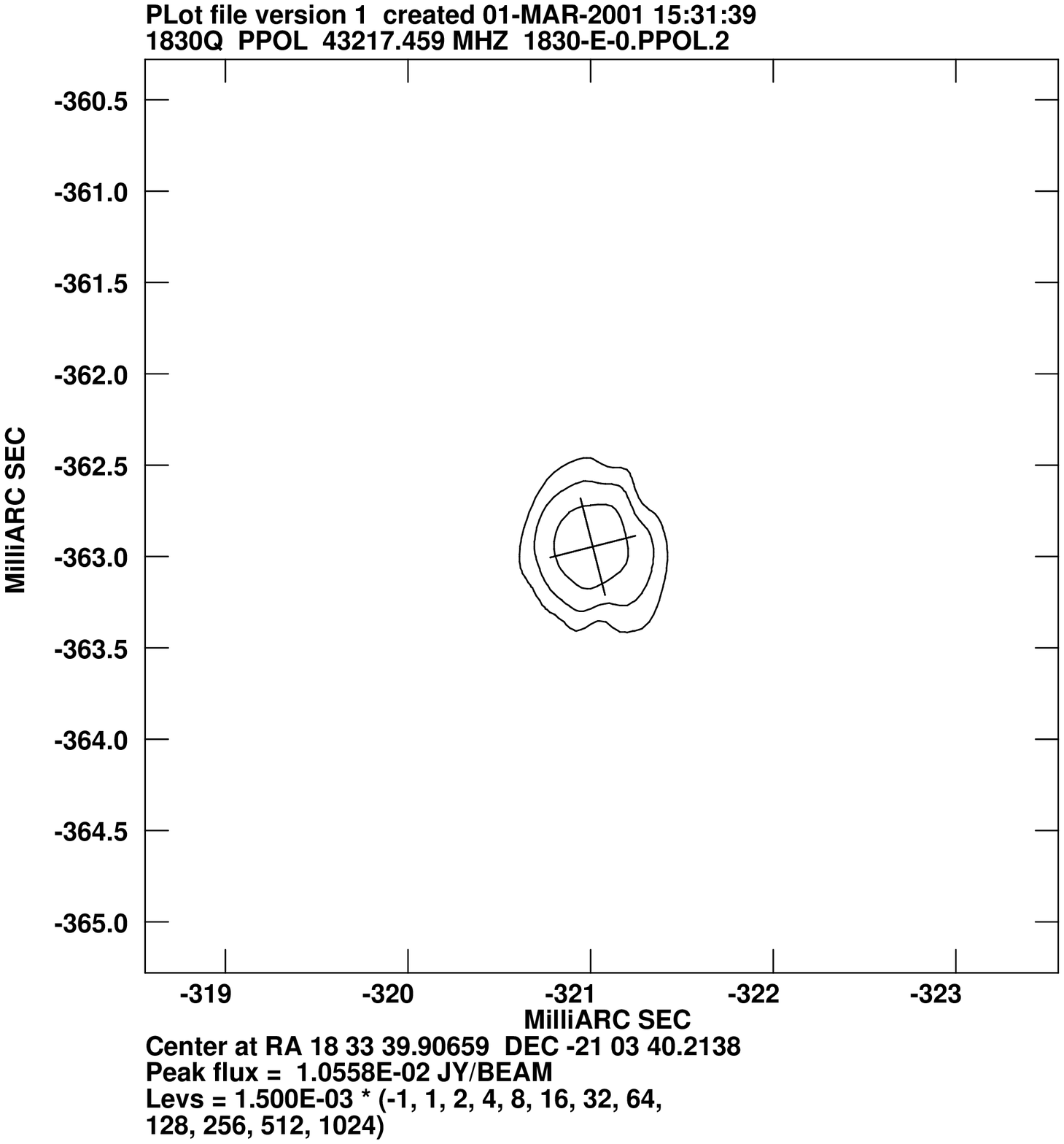}}
\put(140,390){\includegraphics{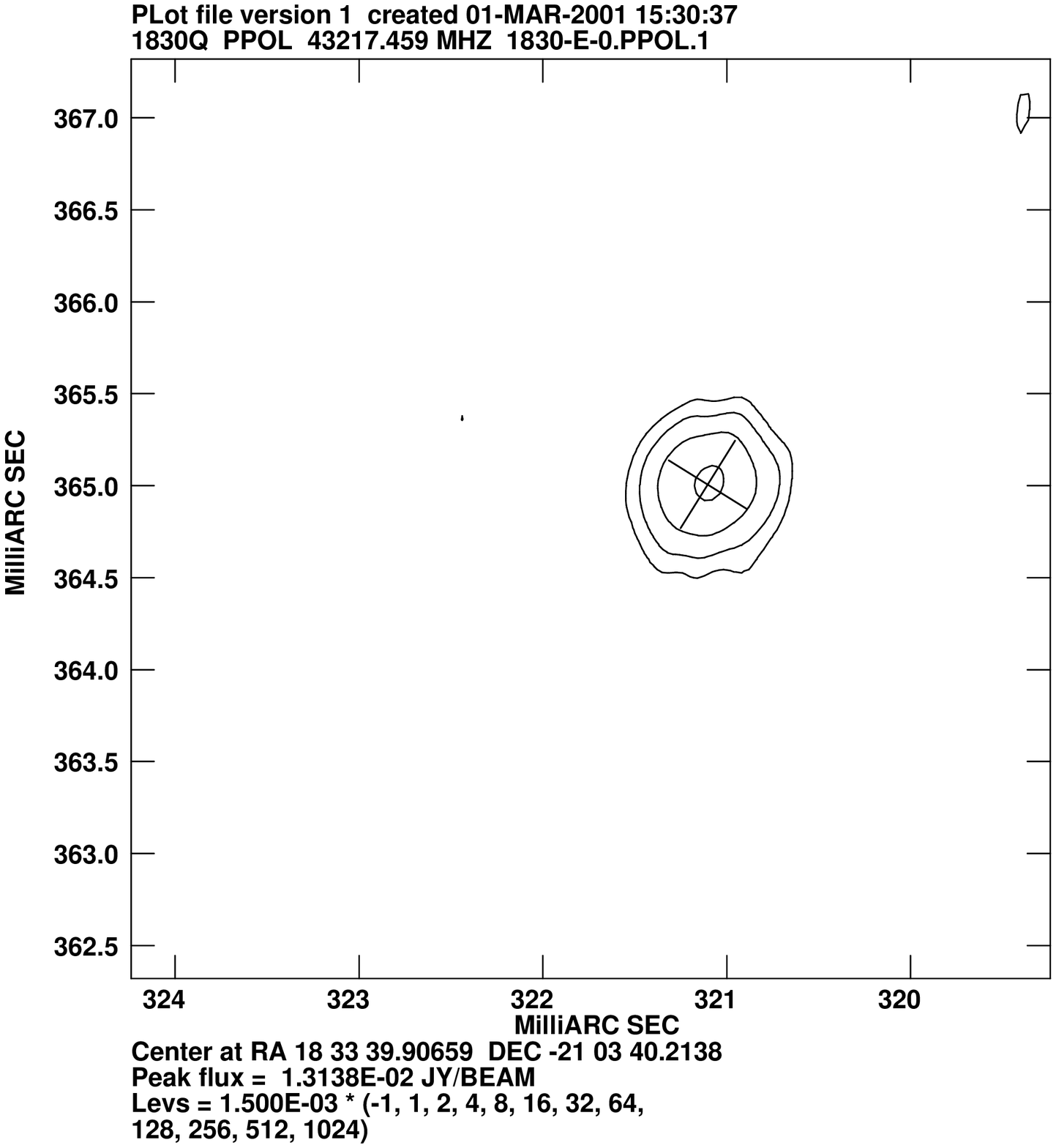}}
\end{picture}
\caption{Polarised intensity contour maps of the SW and NE images for all
eight epochs. The epoch of observation increases from top to bottom,
left to right. Contours are spaced by factors of two in brightness, with
the lowest at five times the r.m.s. noise of 0.3 mJy per beam. All
maps are on the same scale. The FWHM of the circular restoring beam is
0.5 mas. The crosses superimposed on the maps represent the position
and extent of the fitted Gaussian components. The measured time-delay of
$26^{+4}_{-5}$ days implies that structure in the NE image at one
epoch corresponds to that in the SW image roughly two epochs later. }
\label{fig11}
\end{figure*}

\begin{figure*}
\vspace{10cm}
\begin{picture}(40,40)
\put(-200,340){\includegraphics{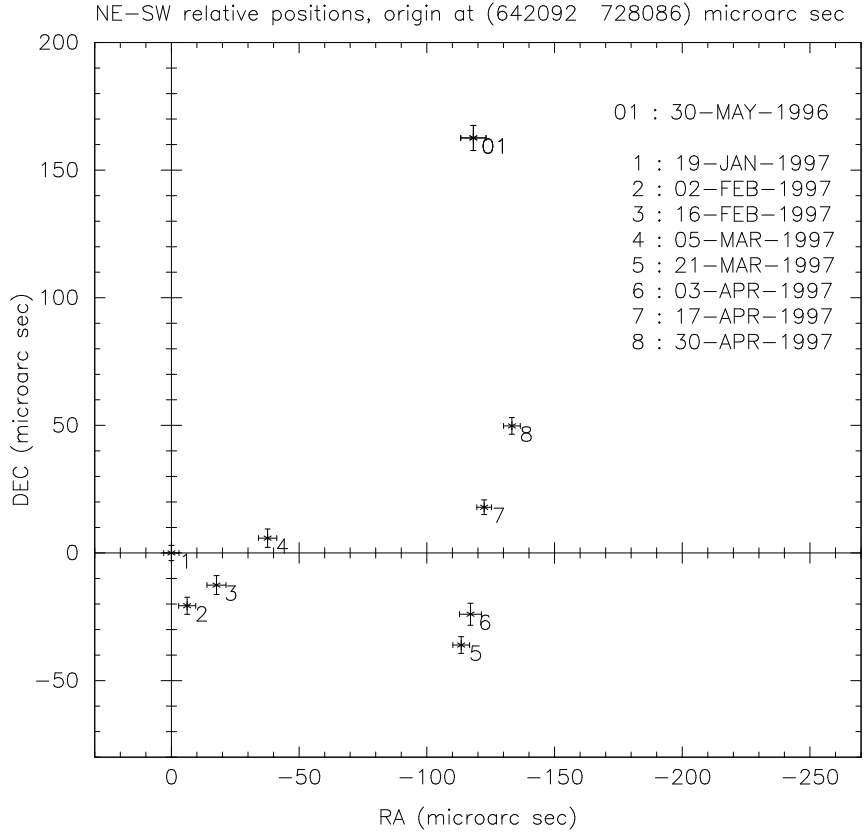}}
\end{picture}
\caption{Deviations in the angular separation of the NE and SW core
  images. The nominal origin of these deviations is defined to be
  $+642071.312 \mu$as and $+727990.0 \mu$as. The numbers associated
  with each of the deviations (see crossed symbols) indicates the
  epoch of the observation.}
\label{fig2}
\end{figure*}

The naturally weighted contour maps of total and \\
polarised intensity
are shown in Fig.~\ref{fig1} and Fig.~\ref{fig11}. The \\
FWHM of the circular restoring beam is 0.5 mas. The \\
off-source rms noise level is 
about 0.8 mJy beam$^{-1}$ in total intensity, and 0.3 mJy beam$^{-1}$ 
in polarised intensity maps at all eight epochs. 
A bright, compact ``core'' and a fainter more extended ``jet'' are clearly
seen on both of the NE \& SW images at all epochs. 
Each of
these core and jet components was fitted by a single elliptical Gaussian
component using {\sc AIPS} task {\sc IMFIT} 
except for the much weaker ``jet'' on the NE image at the 6th epoch, where 
the low flux density prevents reasonable model-fitting. 
One elliptical Gaussian
component was used to fit the single component observed in the
polarised maps of both the NE \& SW images. The results are presented in 
Table~\ref{table1}.

\begin{table*}
\centering
\caption{Total flux densities (in mJy) of the Gaussian fitted components on the 
images. From left to right, we present the total flux intensiy of the core, 
jet and polarised core for the NE and SW images.}
\label{table1}
\begin{tabular}{cccccccccc}

  epoch & NE I-core  & NE I-jet & NE P-core & SW I-core & SW I-jet & SW P-core\\  
  1997 Jan 19 & $505.7 \pm 2.4$ & $72.6 \pm 3.7$ & $13.2 \pm 0.4$ & $428.2 \pm 2.2$ & $57.4 \pm 3.0$ & $11.2 \pm 0.3$ \\
  1997 Feb 02 & $485.1 \pm 2.5$ & $64.2 \pm 3.9$ & $14.3 \pm 0.4$ & $386.1 \pm 2.2$ & $73.4 \pm 3.4$ & $12.9 \pm 0.3$ \\
  1997 Feb 16 & $415.3 \pm 2.5$ & $54.7 \pm 3.9$ & $11.9 \pm 0.3$ & $353.3 \pm 2.2$ & $53.6 \pm 3.2$ & $12.3 \pm 0.4$ \\
  1997 Mar 05 & $478.1 \pm 2.6$ & $59.9 \pm 3.7$ & $15.2 \pm 0.4$ & $406.0 \pm 2.2$ & $62.3 \pm 3.4$ & $14.1 \pm 0.4$ \\
  1997 Mar 21 & $449.8 \pm 2.4$ & $49.2 \pm 3.8$ & $14.2 \pm 0.3$ & $348.6 \pm 2.1$ & $36.7 \pm 2.8$ & $10.5 \pm 0.3$ \\
  1997 Apr 03 & $385.9 \pm 2.5$ & $  -- \pm 2.0$ & $11.6 \pm 0.3$ & $291.9 \pm 2.2$ & $30.2 \pm 2.9$ & $ 9.8 \pm 0.3$ \\
  1997 Apr 17 & $569.0 \pm 2.5$ & $53.4 \pm 4.1$ & $13.6 \pm 0.3$ & $447.3 \pm 2.2$ & $42.9 \pm 3.1$ & $19.3 \pm 0.3$ \\
  1997 Apr 30 & $511.5 \pm 2.6$ & $51.6 \pm 5.5$ & $ 9.7 \pm 0.3$ & $459.9 \pm 2.3$ & $44.3 \pm 3.0$ & $14.5 \pm 0.3$ \\

\end{tabular}
\end{table*}

\section{Results and Discussion}

\subsection{The angular separation between the NE and SW images}

After fitting Gaussians to the NE and SW core components, we measured
the angular separation between the centroids \\of the core radio
emission. We take this as an indicator of the separation between the two
images, since the core is well defined across all epochs. And the core 
is much brighter, thus the errors on the positions are much smaller. 
The uncertainty of the positions of the fitted Gaussian components 
are estimated by the following formula:\\
$
\Delta X \sim {\Theta (beam) \over 2} \cdot {1 \over SNR} \cdot
\sqrt{1 + ({\Theta (imagesize) \over \Theta (beam)})^2}
$

\noindent where SNR is the signal-to-noise ratio of the component, which is determined by 
dividing the peak flux density of the fitted Gaussian component by the 
post-fit r.m.s. error \footnote{The post-fit r.m.s. error is derived from
the difference between the Gaussian model and the data, for the pixels in
the lensed image.} \\
associated with the pixels in the image.
Notice that the \\aberration effect, which is mainly caused by the earth's 
\\orbital motion around the sun, will cause an appararent position shift 
of a source. The amount of the shift(aberration) is a function of the 
position of the source and the time of a year when the source is observed.
Considering two fixed sources on the sky, this will make the measured apparent
separation between the two sources different from the real separation. 
And further, the difference will change at \\different time of a year.
In our case, for the NE \& SW \\images of the gravitational lens system, this  
instantaneous \\differential aberration  
changes by up to several tens $\mu$as across the epochs, much larger
than the positional accuracy of our measurements. In order to compare the 
separations across the epochs, we have to take these effects into account. 
\\We subtracted the aberrations of the NE \& SW images from the apparent 
separation at each epoch. Thus got the real separations between the two 
lensed images at each epoch. The final results are presented in 
Fig.~\ref{fig2}. In the figure, we defined a nominal separation between 
the core images to be that measured at the first epoch of observations, 
viz., $+642092 \mu$as (RA) and $+728086 \mu$as (Dec), relative to the 
SW core. We plot deviations from this nominal separation (also relative 
to the SW core) for each epoch. We have also included the previous 
observation (01:30-MAY-1996) in this figure.  
It can be clearly seen that the relative separation \\between the images of
the core changes with 
epoch. The largest overall deviation of up to 142 $\mu$as 
occurs at the 8th epoch (1997 April 30). The largest 
deviation measured \\between successive epochs is 87 $\mu$as 
(epochs 4 and 5).

A comparison of the nominal separation of the NE and SW core images
with previous 43~GHz VLBA observation \cite{b6} made 
$8$
months
earlier show even larger deviations of 201 $\mu$as. 

We have considered several possible {\it extrinsic\/} \\
explanations for this observed change in the angular \\
separation of the images.  These
include: induced shifts in the centroids of the core brightness
distribution due to \\
milli-lensing by massive ($10^3 - 10^4 M\odot$)
compact objects in the halo of the lens \cite{b20}; and the relative
proper motion between the NE and SW images caused by the transverse
velocity of the lens galaxy across the sky \cite{b22}, though the image
separation does not necesarily change with changing source position. 
However, neither of these effects can reproduce the magnitude 
of the measured deviations
on either the \\shortest (a few weeks) or longest time scales (11
months) considered here.  In addition, the effect of ``image wander''-
an apparent shift in the position of a source caused by \\Galactic
scattering \cite{b12}, also seems to be an unlikely explanation for the
changing image separations at this frequency. In particular, the 
SW deconvolved core size of $\sim 0.6 \times 0.2 $~mas at 1.3 cm \cite{b13} 
and our \\measurement of $\sim 0.228 \times 0.148 $~mas at 7 mm 
scales \\almost linearly with $\lambda$ as expected from simple models of  
synchrotron radio emission. A $\lambda^2$ law would be expected in
the case of ISS (indeed this is observed at longer cm wavelengths 
\cite{b13}). 
This suggests that the \\measured
source size is dominated by its internal radio structure, and that any
interstellar scattering effects are rather weak at $\lambda$~7~mm.

We believe that the most likely explanation for the changing image
separation involves continuous changes in the brightness distribution
of the background radio source on relatively short-time scales
(considerably shorter than that usually probed by VLBI). As the
centroid of the \\
radio emission of the background radio source moves
within the source plane, the separation of the images changes. In \\
addition, the position of one image lags behind the other (the motions
being separated by the time-delay). The changes in the measured angular
separation are a combination of both effects. 

For a FRW universe ($\Omega_{o}=0.3$, $\lambda_{o}=0.7$, H$_{0}=70~km
s^{-1}~Mpc^{-1}$), the shift of $\approx$ 80 $\mu$as (the largest shift \\
between successive epochs separated by 16 days) \\
corresponds to a linear
distance of $\approx$ 0.45~pc at $z_s = 2.507$. Even if one takes into
account the magnification provided by the lens (a factor of $\sim 10$
according to \cite{b7}\cite{b8}), a shift of 80 $\mu$ scales to
$\approx$ 0.14 pc, thus implying (unlensed) superluminal \\velocities in
the rest frame of the background source in \\excess of 10~$c$.  
Given the blazar nature of the background source, and the apparently chaotic
structure surrounding the cores at 43 GHz \cite{b6} which implies the radio jet
in 1830-211 is well beamed towards us, the measured superluminal velocity 
of 10~$c$ might be expected. Another EGRET source 1156+295 \cite{b5}, 
which is probably similar to the background source of 1830, also showed 
superluminal velocity of several c \cite{b30}, but across 10 epochs spanning
a period of 8 years. The detection of source evolution in 1830 on these short 
time-scale of several weeks would be  
impossible if it were not for the fact that this is a lensed system
which provides us with a magnified view and closely spaced multiple
images that allow accurate relative position measurements to be made.

\begin{figure}
\vspace{5cm}
\begin{picture}(40,40)
\put(0,170){\includegraphics{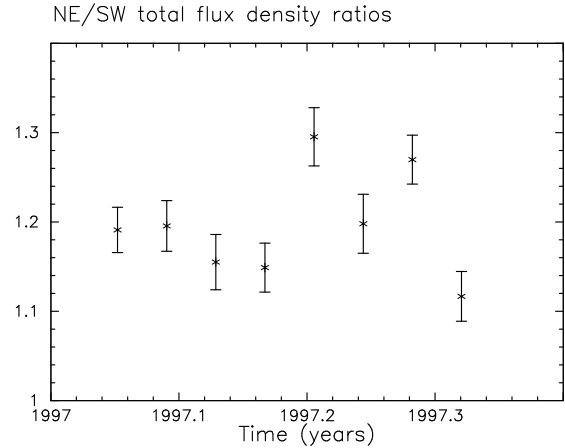}}
\end{picture}
\caption{The flux density ratio between NE and SW image as a function of
time.}
\label{fig3}
\end{figure}

\subsection{The relative magnification matrix}

\subsubsection{Determination of the relative magnification matrix}

Based on flux density ratio arguments, we can directly \\
relate the
``core'' and the ``jet'' in one image to that in the other image. In
addition, we also note that both images are also weakly polarised, with
the position of the polarised \\
emission being significantly offset from
the position of the total intensity core emission. We therefore
identify two \\
vectors in each lensed image: the vector defined by the
core-jet structure and the vector defined by the positional offset of
the polarised emission from the total intensity core \\
position. These
vectors were used to determine a magnification matrix, $M_{SW2NE}$,
that relates the brightness distribution of the SW image to that
observed in the NE image.

\begin{figure*}
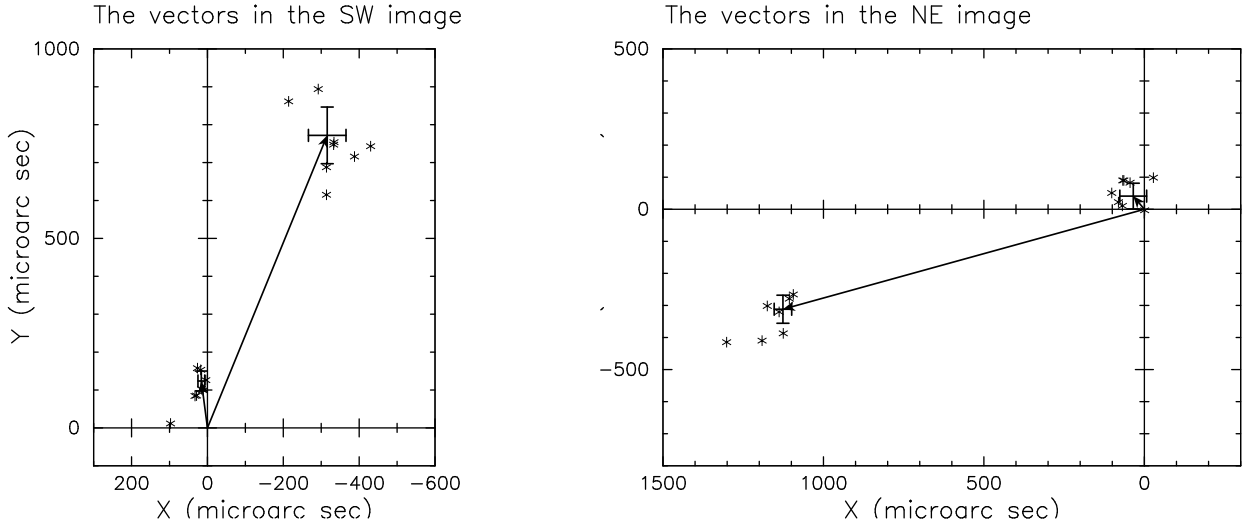

\vspace{4.0cm}
\begin{picture}(200,110)
\put(-180,210){\includegraphics{./swvec.ps}}
\put(80,210){\includegraphics{./nevec.ps}}
\end{picture}
\caption{The two set of vectors in the NE(right) and SW(left) images.
The core-jet separations have larger amplitudes compared with the offsets
between the centroids of the polarised and total intensity emission.}
\label{fig4}
\end{figure*}

Figure~\ref{fig4} shows the vectors corresponding to the positions of
the ``jet'' and the polarised peak with respect to the ``core''. Both
of the vectors vary through the 8 epochs of \\
observation but the
changes are not obviously systematic ({\it e.g.} there is no clear, linear
proper motion of the secondary component with respect to the core). We
have determined the error-weighted averages for these quantities over
all \\
epochs of observation. Since there is a measured time-delay of
$26^{+4}_{-5}$ \cite{b10} days, we have exclude the first two epochs of the
SW image and the last two epochs of the NE image in the averaging process.
The uncertainty of each \\
averaged vector is estimated from the scattering
of the \\
vector through the epochs. 

Using these data we estimate $M_{SW2NE}$:
 
\[M_{(SW2NE)} = \left| \begin{array}{cc}
-2.20 \pm 0.50 &  0.56 \pm 0.19 \\
 1.38 \pm 0.43 &  0.16 \pm 0.17 \end{array} \right|.\]

The errors (1$\sigma$) are estimated from Monte Carlo simulations. 
The determinant
of the matrix is: $-1.13 \pm 0.61$ (1$\sigma$). 
The sign of the determinant indicates 
the two images have opposite parities as predicted by lens models of the 
system. The magnitude of determinant is consistent with
Lovell \\et al.'s \shortcite{b10} measurement for the flux density ratio 
(NE:SW) of $1.52 \pm 0.005$, and our own measurements through the 8 
epochs of observation, presented in Fig~\ref{fig3}. 
The ratio (NE:SW) changes with epoch, ranging from 1.1 to 1.4, in good
agreement with the determinant of the matrix \\
(determined purely from
positional information only). The changes of the ratio are most likely
due to intrinsic brightness changes in the background radio source
combined with the time-delay between the images (changes in the SW \\
image lag behind the NE image by 3-4 weeks).

\subsubsection{Applying the matrix to previous VLBI observations} 

We have applied our matrix to previous VLBA 15~GHz \\
observations
of PKS~1830-211 \cite{b6}. We find that the peak of the
faint extended component that lies \\approximately (-5.0,5.0)mas to 
the north-west of the 15~GHz SW core, maps to (13.8, -6.1)mas to
the sourth-east of the NE core in the NE image with an uncertainty 
of 2.2 mas(1$\sigma$). This overlaps with the extended component that 
lies \\approximately 13 mas to the south-east of the 15~GHz NE core.
Given that the positions of these faint
components are not very well defined (due to their extended nature),
this close correspondence strongly suggests that these features are
associated, multiply-lensed emission.

Our new matrix {\it cannot\/}, however, explain the fact that the
prominent 10-mas-long jet, clearly seen emerging from the core of the NE
image at 15~GHz in a north-westerly \\direction, has no obvious
counterpart in the SW image. The matrix predicts that the position of
the hot-spot at the tip of the jet, approximately 10 mas to the
north-west of the NE core, should map on to a region located 5 mas to
the south-east of the SW core with an uncertainty of tens of mas(1$\sigma$). 
This large positional uncertainty makes it difficult to identify the 
jet's counterpart in the SW image. Even if we ignore this large uncertainty,    
such emission, including the rest of the 
extended jet, should have been easily detectable at 15~GHz and 8.4~GHz. 
Nair \& Garrett \shortcite{b14} have recently \\
attempted to explain the
apparently singular jet in the NE image as a perturbation in the lens
potential probed by that image. In their model Nair and Garrett argue 
that this perturbation arises from sub-structure in the lens e.g. 
the chance alignment of the NE radio image with a massive discrete
structure, such as a globular cluster in the lens galaxy.

\subsection{The flux ratios}

Since the VLBA observations at 43~GHz were severely \\
affected by the atmospheric opacity effect, so the absolute flux 
density scale in Table~\ref{table1} varied between one 
epoch and another. Thus it is not possible to get
meaningful light curves from the fluxes presented in Table~\ref{table1}. 
However, \\taking the data from Table~\ref{table1}, we have considered 
various flux density
ratios in the two images. The core-jet flux \\density ratio and 
the fractional polarisation of the two \\images change with epoch, but 
neither of these two ratios show obvious correlation after the time-delay
of $24^{+5}_{-4}$ days has been compensated for.   
The jet components in the two images are normally more extended, thus making 
their flux density estimates less
accurate than that of the core components.
If the source has detailed polarisation structure that can not be 
resolved by the VLBI array, the polarisation flux density  measurement
will also have large errors. 
In the light of the above, the results obtained are not surprising.

\section{Summary} 

We have presented eight-epoch, 43~GHz, dual-polarisation VLBA
observations of the gravitational lens system, PKS 1830-211.
``Core-jet'' structure was
clearly seen in both lensed images at all epochs. 
The
position of the  polarised intensity emission is observed to be offset
from the peak of total intensity maps. The offset is about one third of
the synthesised beam and is observed at all eight epochs and in both
images.
 
The relative separation between the ``core'' centroid positions of the 
two lensed images changes by up to 87 $\mu$as between subsequent epochs. A 
comparison of these recent \\
separation measurements with previous 43~GHz \\
observations \cite{b6} shows even larger deviations of up to 201 
$\mu$as over a period of 27 weeks. The measured changes
are most likely produced by evolution in the brightness distribution of
the background source, enhanced by the magnification of the lens.

The multi-component core-jet radio structure observed in both images,
together with the spatial offset between the polarised and total
intensity core emission, permit us to \\
estimate a relative magnification
matrix for this system. 
The determinant of the matrix is $-1.13 \pm 0.61$, in good agreement 
with the flux density ratio of the images.
The matrix presented here can be an important input for the
construction of realistic lens models for this system and thus
associated estimates of $H_0$. 

Application of this matrix to previous VLBI observations meets with
mixed success. In one case it is reasonably successful in relating a
faint extended component observed in each of the NE and SW images at
15~GHz.  However, it also predicts that the extremely prominent
10-mas-long jet associated with the 15~GHz NE image, should be easily
detected in the SW image. The fact that the high quality 15 GHz and 8.4
GHz VLBA observations 
(Garrett et al. 1997, Guirado et al. 1999) 
do not detect such a jet, provides strong evidence for sub-structure
in the lens. More realistic, multi-component lens mass distributions 
probably need to be invoked ({\it e.g.} Nair \& Garrett \shortcite{b14}).

\section*{Acknowledgments}

This research was supported by the grant for collaborative research in
radio astronomy of the Royal Dutch and Chinese Academies of Science
(KNAW and CAS) and was also partially supported by a grant from the NSFC.
The National Radio Astronomy Observatory is a facility of the National
Science Foundation operated under cooperative agreement \\
by Associated Universities, Inc. C. Jin wishes to thank JIVE staff, 
in particular R.T.
Schilizzi, L. Sjouwerman, H.J. van Langevelde and L.I. Gurvits for valuable
discussions on \\various aspects of VLBI data analysis. S. Nair thanks Mark
Walker for useful comments.

\label{lastpage}

\end{document}